\documentclass{article}

\usepackage{arxiv}

\usepackage{amssymb}

\usepackage{graphicx}
\usepackage[space]{grffile}
\usepackage{latexsym}
\usepackage{textcomp}
\usepackage{longtable}
\usepackage{tabulary}
\usepackage{booktabs,array,multirow}
\usepackage{amsfonts,amsmath,amssymb}
\usepackage{url}
\usepackage{hyperref}
\hypersetup{colorlinks=true,linkcolor=red,citecolor=black,urlcolor=magenta}
\usepackage{etoolbox}
\usepackage{natbib}
\usepackage{setspace}
\usepackage[title]{appendix}

\newif\iflatexml\latexmlfalse

\AtBeginDocument{\DeclareGraphicsExtensions{.pdf,.PDF,.eps,.EPS,.png,.PNG,.tif,.TIF,.jpg,.JPG,.jpeg,.JPEG}}

\usepackage[utf8]{inputenc}
\usepackage[english]{babel}

\usepackage{siunitx}
\usepackage[T1]{fontenc}    
\usepackage{bm}
\usepackage{mathtools}
\usepackage{nicefrac}       
\usepackage{microtype}      
\usepackage{lipsum}
\usepackage[ruled,vlined]{algorithm2e}
\usepackage{xcolor}
\usepackage{afterpage}
\usepackage{pdflscape}
\usepackage{placeins}
\usepackage[normalem]{ulem}

\usepackage{wasysym}

\usepackage{framed}
\usepackage[framemethod=tikz]{mdframed}%

\usepackage[textsize=tiny]{todonotes}
\definecolor{org}{rgb}{1,0.53,0.0}

\definecolor{mrg}{rgb}{0.1,0.45,0.1}

\newtheorem{example}{Example}

\makeatletter
\newcommand\makebig[2]{%
  \@xp\newcommand\@xp*\csname#1\endcsname{\bBigg@{#2}}%
  \@xp\newcommand\@xp*\csname#1l\endcsname{\@xp\mathopen\csname#1\endcsname}%
  \@xp\newcommand\@xp*\csname#1r\endcsname{\@xp\mathclose\csname#1\endcsname}%
}
\makeatother
\makebig{biggg} {3.0}
\makebig{Biggg} {3.5}
\makebig{bigggg}{4.0}
\makebig{Bigggg}{4.5}

\newcommand{\w}{\bm{w}}
\newcommand{\x}{\bm{x}}
\newcommand{\y}{\bm{y}}
\newcommand{\z}{\bm{z}}

\newcommand{\X}{\mathbf{X}}
\newcommand{\Y}{\mathbf{Y}}
\newcommand{\Z}{\mathbf{Z}}
\newcommand{\W}{\mathbf{W}}

\definecolor{blue_custom}{HTML}{0485d1}

\newcommand{\T}{\bm{T}}

\newcommand{\SKR}{\bm{S}} 

\newcommand{\R}{\mathbb{R}}

\title{Ensemble transport smoothing\\ Part II: Nonlinear updates}
\renewcommand{\shorttitle}{Ensemble transport smoothing, Part II}

\author{Maximilian Ramgraber \\
	Department of Aeronautics and Astronautics\\
	Massachusetts Institute of Technology\\
	Cambridge, MA 02139 \\
	\texttt{mramgrab@mit.edu} \\
	\And
	Ricardo Baptista \\
	Department of Aeronautics and Astronautics\\
	Massachusetts Institute of Technology\\
	Cambridge, MA 02139 \\
	\texttt{rsb@mit.edu} \\
	\And
    Dennis McLaughlin  \\
    Department of Civil and Environmental Engineering\\
    Massachusetts Institute of Technology\\
    Cambridge, MA 02139 \\
    \texttt{dennism@mit.edu} \\
	\And
	Youssef Marzouk \\
	Department of Aeronautics and Astronautics\\
	Massachusetts Institute of Technology\\
	Cambridge, MA 02139 \\
	\texttt{ymarz@mit.edu} \\
}

\begin{document}

\maketitle

\begin{abstract}
Smoothing is a specialized form of Bayesian inference for state-space models that characterizes the posterior distribution of a collection of states given an associated sequence of observations. 
\citet{Ramgraber2022underUpdates} proposes a general framework for transport-based ensemble smoothing, which includes linear Kalman-type smoothers as special cases. Here, we build on this foundation to realize and demonstrate nonlinear backward ensemble transport smoothers. We discuss parameterization and regularization of the associated transport maps, and then examine the performance of these smoothers for nonlinear and chaotic dynamical systems that exhibit non-Gaussian behavior. In these settings, our nonlinear transport smoothers yield lower estimation error than conventional linear smoothers and state-of-the-art iterative ensemble Kalman smoothers, for comparable numbers of model evaluations.
\end{abstract}


\section{Introduction}\label{sec:introduction}
Smoothing in the Bayesian setting recursively characterizes the posterior distribution $p(\x_{1:t}|\y_{1:t}^{*})$ of a sequence of states $\x_{1:t}$ given a sequence of observations $\y_{1:t}^{*}$. The most widely used ensemble approaches for smoothing are \textit{sequential Monte Carlo} methods \citep{Doucet2009ALater,Klaas2006FastParticles} and \textit{ensemble Kalman} methods \citep{Asch2016DataApplications,Evensen2003TheImplementation}. Sequential Monte Carlo methods can characterize arbitrary distributions using sequential importance sampling and resampling, but typically require very large sample sizes to mitigate weight collapse \citep{Snyder2008ObstaclesFiltering,Snyder2015PerformanceProposal}. By contrast, ensemble Kalman-type methods avoid the use of weights, but are based on affine prior-to-posterior updates that are consistent only if all distributions involved are Gaussian. In the context of smoothing, such methods include the ensemble Kalman smoother (EnKS) \citep{Evensen2000AnDynamics}, which has inspired numerous algorithmic variations such as the ensemble smoother with multiple data assimilation \citep{Emerick2013EnsembleAssimilation} and the iterative ensemble Kalman smoother (iEnKS) \citep{Bocquet2014AnSmoother,Evensen2019EfficientMatching}, as well as backwards smoothers such as the ensemble Rauch--Tung--Striebel smoother (EnRTSS) \citep{Raanes2016OnSmoother}.

These two classes of methods occupy opposite ends of a spectrum that ranges from an emphasis on statistical generality at one end to an emphasis on computational efficiency at the other. This trade-off complicates design decisions for smoothing problems that are at once non-Gaussian and computationally expensive. 
In such problems, non-Gaussianity renders linear Kalman-type methods generally inconsistent with the true Bayesian solution.
At the same time, the ensemble size required for importance sampling in practical non-Gaussian problems often proves prohibitive. Left with little recourse, practitioners typically resort to Kalman-type methods, sacrificing statistical fidelity for computational feasibility in non-Gaussian systems. Many researchers recognize the limitations of this approach, and substantial effort has been devoted to increasing the robustness of affine-update algorithms to deviations from the Gaussian setting \citep[e.g., ][]{Bocquet2014AnSmoother,Sarkka2010BayesianSmoothing,Schoniger2012ParameterTomography}. Despite these efforts, the fundamental limitations of affine updates remain.

\textit{Transport methods} \citep[e.g.,][]{Marzouk2017SamplingIntroduction,Spantini2018InferenceCouplings,pulido2019sequential,hao2023hybrid} offer 
a new path for progress, as they generalize the sample-efficient prior-to-posterior transformations of Kalman type methods to fully \textit{nonlinear} updates. 
Prior work by \citet{Spantini2022CouplingFiltering} successfully established a nonlinear generalization of the ensemble Kalman filter \citep{Evensen1994SequentialStatistics,Evensen2003TheImplementation}. A companion manuscript to the present paper \citep{Ramgraber2022underUpdates} proposes a general framework for nonlinear \textit{ensemble transport smoothing}. In this framework, different affine-update Kalman algorithms emerge as special cases, including the ensemble Kalman smoother (EnKS) \citep{Evensen2000AnDynamics} and the ensemble Rauch--Tung--Striebel smoother (EnRTSS) \citep[e.g.,][]{Raanes2016OnSmoother}. In the present paper, we demonstrate and explore the use of nonlinear updates in ensemble transport smoothing, 
evaluating the performance of the resulting algorithms in nonlinear and chaotic systems of varying dimensionality, and comparing our results to state-of-the-art \textit{iterative ensemble Kalman smoothers} (iEnKS) \citep{Bocquet2014AnSmoother}. 
We specifically discuss parameterizations of nonlinear transport maps for filtering and smoothing, and derive new localization approaches, based on identifying conditional independence, that enable efficient smoothing through the construction of sparse maps.

This manuscript is structured as follows. We first recall the basics of ensemble transport methods and conditional sampling in Section~\ref{sec:transport_maps}. Readers coming from Part 1 \citep{Ramgraber2022underUpdates} of this two-part manuscript can safely skip this section except for Section~\ref{subsec:monotonicity}, which describes parameterizations for nonlinear monotone transport maps; further details on the relevant optimization procedures are in Appendices~\ref{sec:AppendixB} and \ref{sec:AppendixA}. Section~\ref{sec:ensemble_transport_smoothing} then discusses how to use these maps in various backward smoothing recursions. We examine the performance of the resulting smoothers in Section~\ref{sec:experiments} and discuss our findings in Section~\ref{sec:discussion}. Notation used in this manuscript is summarized in Table~\ref{tab:nomenclature}.

\section{Nomenclature}

\begin{table}[!ht]
    \caption{Notation used in this study.}
    \begin{center}
    \begin{tabular}{ r l }
     $x$ or $S(x)$ & scalar-valued variables or functions \\ 
     $\x$ or $\SKR (\x)$ & vector-valued variables or functions \\  
     $\x$ & state variables \\
     $\y$ & predicted (unrealized) observations \\
     $\y^{*}$ & realized values of observations \\
     $N$ & ensemble size \\
     $\X$ or $\Y$ & ensemble representations of $\x$ or $\y$ (e.g., $\X = \{\x^i\}_{i=1}^N$) \\
     $\w$ or $\W$ & generic random variable and its ensemble representation \\
     $x \sim p$ & $x$ is distributed according to the probability distribution $p$ \\
     $K$ & dimensionality of $\w$ \\
     $M$ & dimensionality of the upper map block \\
     $D$ & dimensionality of the lower map block \\
     $1 \leq s \leq t$ & time step (often in a subscript) \\
     $1 \leq k \leq K$ & vector component index (often in a subscript) \\
     $1 \leq n \leq N$ & ensemble sample index (often in a superscript) \\
     $\X_{t}^{*}$ & $\X_{t}$ conditioned on $\y_{t}^{*}$; i.e., if $\X_{t}\sim p(\x_{t}|\y_{1:t-1}^{*})$, then $\X_{t}^{*}\sim p(\x_{t}|\y_{1:t}^{*})$ \\
     $p$ & target distribution\\
     $\eta$ & reference distribution; usually standard Gaussian $\mathcal{N}(\mathbf{0},\mathbf{I})$ \\
     $\SKR$ & transport map \\
     $\T$ & composite transport map \\
     $\SKR_{\sharp}p$ & pushforward distribution \\
     $\SKR^{\sharp}\eta$ & pullback distribution \\
    \end{tabular}
    \end{center}
    \label{tab:nomenclature}
\end{table}

\section{A brief introduction to transport methods} \label{sec:transport_maps}
In this section, we briefly review the foundations of transportation of measure and introduce the properties that enable conditional sampling and thus general Bayesian inference.
Further details can be found in~\citet{Marzouk2017SamplingIntroduction}, \citet{Spantini2018InferenceCouplings}, and \citet{ElMoselhy2012BayesianMaps}.

\subsection{Change of variables}

Transport methods can characterize probability distributions such as those arising in Bayesian inference problems. Fundamentally, these methods seek an invertible map $\SKR$ that transforms a target random variable $\w$ with a probability density function $p(\w)$ to a reference random vector $\z$ with a probability density 
$\eta(\z)$, such that $\z=\SKR(\w)$ with equality in the distributional sense; see for example~\citet{Villani2007OptimalNew}. Conventionally, the density $p$ is assumed to be known only up to a normalizing constant or through samples, and the reference density $\eta$ is user-specified to be easy to sample from. In this work we take $\eta$ to be the multivariate standard Gaussian density.

For a continuously differentiable map, the target and reference densities are related through the map via the \textit{change-of-variables} formula \citep[e.g.,][]{Spantini2018InferenceCouplings}
\begin{equation}
p(\w) = \SKR^{\sharp}\eta(\w) \coloneqq \eta(\SKR(\w))\det \mathbf{\nabla} \SKR(\w).
\label{eq:change_of_variables}
\end{equation}
If the map is exact, this formula describes how the reference density $\eta$ is transformed into a \textit{pullback density} $\SKR^{\sharp}\eta$ to match the target density $p(\w)$. 
The term $\det \mathbf{\nabla} \SKR(\w)$ accounts for the differential change in volume between the coordinate systems of $\w$ and $\z$, ensuring that maps which stretch or squeeze the coordinate systems preserve probability mass. Since the maps considered here are invertible, an analogous change-of-variables formula may be used to approximate the reference density $\eta(\z)$ at an argument $\z$ via the \textit{pushforward density} $\SKR_{\sharp}p$ that depends on $p(\w)$, evaluated at $\w=\SKR^{-1}(\z)$.

\subsection{Triangular maps}

While there exist  an infinite number of transport maps that can push forward one distribution to another, one convenient structure is provided by the \textit{Knothe--Rosenblatt} (KR) \textit{rearrangement}~\citep{Rosenblatt1952RemarksTransformation}. The KR map $\SKR$ is a lower-triangular function of the form 
\begin{equation}
\SKR(\w)=\begin{bmatrix*}[l]
    S_{1}(w_{1}) \\
    S_{2}(w_{1},w_{2}) \\
    \vdots \\
    S_{K}(w_{1},\dots,w_{K})
\end{bmatrix*} 
=\begin{bmatrix*}[l]
    z_{1} \\
    z_{2} \\
    \vdots \\
    z_{K}
\end{bmatrix*},
\label{eq:decomposed_S}
\end{equation} 
where the $k$th map component function $S_{k}\colon\mathbb{R}^k\rightarrow\mathbb{R}$ depends only on the first $k$ inputs of $\w$ and $S_k(w_1,\dots,w_{k-1},w_{k})$ is monotone increasing in $w_{k}$ for all $(w_1,\dots,w_{k-1}) \in \mathbb{R}^{k-1}$.
These structural properties not only guarantee that $\SKR$ is invertible, but also that the determinant of the map's Jacobian 
in Equation~\ref{eq:change_of_variables} can be evaluated efficiently as the product of its diagonal entries, i.e., $\det \nabla S = \prod_{k=1}^{K} \frac{dS_k}{dw_k}$. If $p$ is fully supported on $\mathbb{R}^K$, the KR rearrangement is also the unique transport map with the form given in Equation~\ref{eq:decomposed_S} that satisfies $S^\sharp\eta = p$~\citep{Bogachev2005TriangularMeasures}.

\subsubsection{Inversion and conditioning}\label{subsec:inv_and_cond}

While the components of a triangular map $\SKR$ can be evaluated independently at an input $\w$, inverting the map relies on solving a sequence of one-dimensional root finding problems. Starting from the top of the map, each map component inversion depends on the inversion of the previous components. Let $S_{k}^{-1}(w_1,\dots,w_{k-1}; \, \cdot \, )$ denote the inverse of the scalar-valued map $w_{k} \mapsto S_{k}(w_1,\dots,w_{k-1},w_{k})$ given $(w_1,\dots,w_{k-1}) \in \mathbb{R}^{k-1}$. 
With this definition of the scalar inverse $S_{k}^{-1}$, the inverse of the multivariate map $\SKR$ in Equation~\ref{eq:decomposed_S} is given by: 
\begin{equation}
\SKR^{-1}(\z)=\begin{bmatrix*}[l]
    S_{1}^{-1}(z_{1}) \\
    S_{2}^{-1}(w_{1}; z_{2}) \\
    \vdots \\
    S_{K}^{-1}(w_{1},\dots,w_{K-1}; z_{K})\end{bmatrix*}=\begin{bmatrix*}[l]
    w_{1} \\
    w_{2} \\
    \vdots \\
    w_{K}
\end{bmatrix*}.
\label{eq:inverse_decomposed_S}
\end{equation}

An important property of triangular maps is that the individual components of $\SKR^{-1}$ can be used to sample from various conditionals of the target density $p$. The process is straightforward if the components of the reference random variable are mutually independent or, equivalently, if the reference density $\eta(\z)$ can be written as the product of its scalar marginals, i.e., $\eta(\z) = \prod_{k=1}^{K} \eta_k(z_k)$. In this case, samples from the marginal conditionals of $p$ are obtained from the map component inverses as follows: 
\begin{equation}
\begin{aligned}
w_{1}^{i} &= S_{1}^{-1}(z_{1}^{i}) &&\sim p(w_{1}) && \\
w_{2}^{i} &= S_{2}^{-1}(w_{1}^{i};z_{2}^{i}) &&\sim p(w_{2}|w_{1}) \\
w_{3}^{i} &= S_{3}^{-1}(w_{1}^{i},w_{2}^{i};z_{3}^{i}) &&\sim p(w_{3}|w_{1},w_{2}) \\
&\;\;\vdots && \qquad\vdots\\
w_{K}^{i} &= S_{K}^{-1}(w_{1}^{i},\dots,w_{K-1}^{i};z_{K}^{i}) &&\sim p(w_{K}|w_{1},\dots,w_{K-1}),
\end{aligned}
\label{eq:marginal_conditional}
\end{equation}
given $z_k^i \sim \eta_k$ for all $k$. In this example, the set of all the individual $w_{k}^{i}$ samples follows a joint density $p(\w_{1:K})$ that can be constructed from a product of the corresponding $K$ marginal conditionals \citep[e.g.,][]{Park2017}:
\begin{equation}
p(\w_{1:K}) = p(w_{1})p(w_{2}|w_{1})p(w_{3}|w_{1},w_{2})\cdots p(w_{K}|w_{1},\dots,w_{K-1}).
\label{eq:marginal_conditional_telescopic}
\end{equation} 
In order to exploit this important factorization property, we assume in the remainder of this paper that the components of the random reference variable $\z$ are mutually independent. In our computational examples, these independent reference components follow a multivariate standard Gaussian probability distribution.

Equation~\ref{eq:marginal_conditional_telescopic} reveals three useful properties of transport maps \citep{Spantini2018InferenceCouplings}. First, the factorization of $p(\w_{1:K})$ in Equation~\ref{eq:marginal_conditional_telescopic} depends on an ordering of the variables $\w_{1:K}$ that is selected when constructing $\SKR$. We are free to choose this ordering to facilitate the solution of a particular inference problem. Ordering can also affect the necessary map complexity.

Second, we can exploit conditional independence by removing variable dependencies from the transport map components $S_{k}$. For instance, if $w_3$ is conditionally independent of $w_2$ given $w_1$ (which is written as $w_3\perp \!\!\! \perp  w_2 \; | \; w_1$) we have $p(w_3|w_1,w_2) = p(w_3|w_1)$, and $S_{3}^{-1}(w_1;z_3)$ for $z_3 \sim \eta_3$ samples exactly from the marginal conditional $p(w_3|w_1)$.

Finally, triangular maps can characterize conditionals of the joint distribution $p(\w_{1:K})$, such as the posterior distributions in Bayesian inference. Replacing the output of the map component inverses $S_{j}^{-1}$ in Equation~\ref{eq:inverse_decomposed_S} with values $w_{j}^{*}$ for $1 \leq j < k$  yields samples from the conditional density $p(\w_{k:K}|\w_{1:k-1}^{*})$.

\begin{figure}
  \centering
  \includegraphics[width=\textwidth]{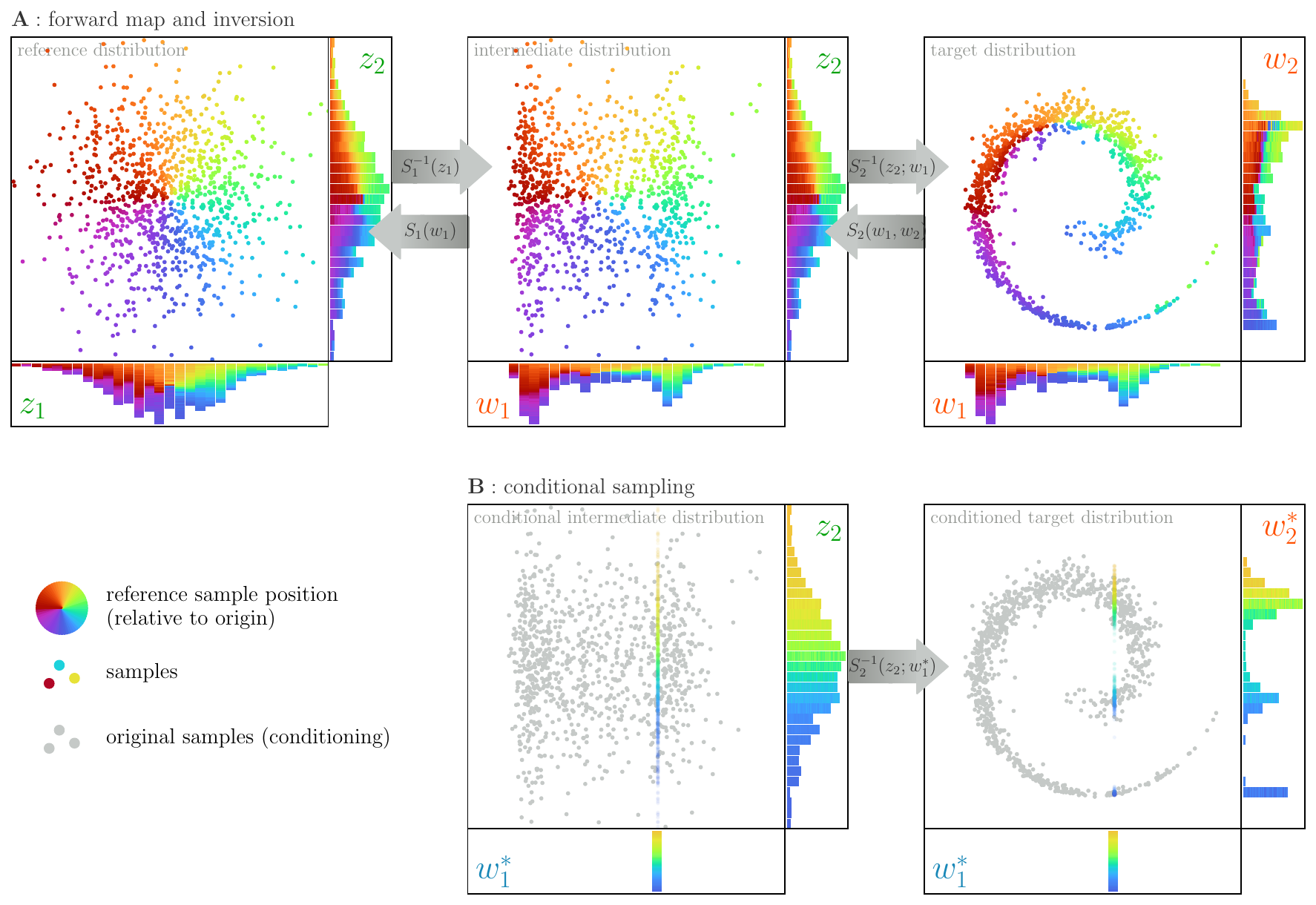}
  \caption{An example of a nonlinear triangular transport mapping between a Gaussian reference distribution $\eta(\z)$ and a non-Gaussian target distribution $p(\w)$. Samples from the two-dimensional joint distributions are illustrated by points, and
  marginal or conditional distributions are illustrated by histograms along the variables axes. (A) The forward map $\SKR(\w)$ (top row, from right to left) and its inverse $\SKR^{-1}(\z)$ (top row, from left to right) operate via an implicit intermediate distribution (top row, center). Each scalar map component, i.e., $S_{1}^{-1}(z_1)$ or $S_{1}(w_1)$, transforms a single variable. (B) Slicing the intermediate distribution at an observation $w_{1}^{*}$  (bottom row, left) generates a set of samples (represented by $w_{2}^{*}$) from the conditional  distribution $p(w_{2}|w_{1}^{*})$ (bottom row, right). This conditional distribution is illustrated by the multi-modal histogram along the $w_{2}^{*}$ axis}
  \label{fig:conditioning}
\end{figure}

Figure~\ref{fig:conditioning} illustrates conditional sampling using triangular maps in greater detail. Consider first the forward map (Figure~\ref{fig:conditioning}A, right to left) that maps the target random variable $(w_1,w_2)$ to reference samples. For the inverse (Figure~\ref{fig:conditioning}A, left to right), we start with samples $\z=(z_{1},z_{2})$ from the reference distribution $\eta$ (left). We begin (Equation~\ref{eq:inverse_decomposed_S}) by inverting the first map component $S_{1}^{-1}(z_{1})$. This yields samples $(w_{1},z_{2})$ from an intermediate distribution (center), which has marginals $p(w_{1})$ from the target and $\eta(z_2)$ from the reference. Samples from this intermediate distribution serve as input for the inverse of the second map component $S_{2}^{-1}(z_{2};w_{1})$, which transforms the second marginal and  yields samples $\w=(w_{1},w_{2})$ from the target distribution $p$ (right).

The conditioning operation in Figure~\ref{fig:conditioning}B uses this inversion process to sample \textit{specific} conditionals of the target distribution. This is achieved by skipping the first map component's inversion and replacing its original output $w_{1}$ with user-specified conditioning values $w_{1}^{*}$, on which we wish to condition (center). Inverting the second map component  $S_{2}^{-1}(z_{2};w_{1}^{*})$ with this specified input inserted yields samples $w_{2}^{*}$ from the conditional $p(w_{2}|w_{1}^{*})$ (right).
We use this capability of conditional sampling within the smoothing algorithms discussed in Section~\ref{sec:ensemble_transport_smoothing}. 

This conditioning process is particularly important for Bayesian inference. To illustrate this, let $w_{1}$ correspond to an observation and $w_{2} = x$ correspond to an unknown state. In this case we can define the (joint) target pdf $p(w_{1},w_{2})$ as the product between the prior $p(w_{2})$ and a likelihood function for  $p(w_{1}|w_{2})$. By ordering the variables as $(w_{1},w_{2})$, the  density factorizes according to Equation~\ref{eq:marginal_conditional_telescopic}. We can then follow the procedure outlined in Figure~\ref{fig:conditioning}B to generate samples from the posterior density $p(w_{2}|w_{1}^{*})$ by inserting a conditioning value $w_{1}^{*}$ where we would otherwise carry out a scalar inversion. We use this capability in the smoothing algorithms discussed in Section~\ref{sec:ensemble_transport_smoothing}.

\subsection{Parameterization of monotone maps}\label{subsec:monotonicity}
For triangular maps, monotonicity can be ensured by parameterizing each map component $S_{k}(\w_{1:k})$ in Equation~\ref{eq:decomposed_S} to be monotone by construction in its last argument $w_{k}$ \citep{Marzouk2017SamplingIntroduction,Spantini2022CouplingFiltering}. In this study, we consider two different approaches to ensure this condition is met: \textit{integrated maps} in Section~\ref{subsubsec:integrated} and \textit{separable maps} in Section~\ref{subsubsec:seperable_maps}.

\subsubsection{Monotonicity via integrated maps}\label{subsubsec:integrated}

A flexible formulation that does not restrict the form of each component beyond being continuously differentiable is to enforce monotonicity through a \textit{rectifier} and \textit{integration}. The integrated maps represent each component as 
\begin{equation}
    S_{k}(\w_{1:k}) = g(\w_{1:k-1}) + \int_{0}^{w_{k}} r (h(\w_{1:k-1},\omega)) d \omega,
    \label{eq:integrated_rectifier}
\end{equation}
where $h \colon \R^{k} \rightarrow \R$ is a (possibly non-monotone) function which may depend on all arguments $\w_{1:k}$, and $g\colon \R^{k-1} \rightarrow \R$ is a function which depends on only the first $k-1$ arguments $\w_{1:k-1}$ \citep{Marzouk2017SamplingIntroduction}. Since $h$ is not in general monotone with respect to $w_{k}$, 
Equation~\ref{eq:integrated_rectifier} applies a \textit{rectifier}, i.e., a strictly positive function $r\colon \mathbb{R} \rightarrow \mathbb{R}_{+}$, to $h$ and integrates the resulting function over $w_k$. This process ensures that $\partial_{w_k} S_{k}(\w_{1:k}) = r(h(\w_{1:k})) > 0$, and hence $S_k$ is monotone in $w_k$. 
Examples of rectifiers that have been shown to yield favorable properties for learning the map include the 
softplus function $r(z) = \log(1+ \exp(z))$ and the shifted exponential linear unit~\citep{Baptista2020OnMaps}. 
\begin{example}
To provide further intuition, consider a concrete example using the following linear expansion of $g$ and $h$ in terms of low-degree polynomials
\begin{equation}
    \begin{aligned}
        &g(w_{1},w_{2}) &&= c_{0} + c_{1}w_{1} + c_{2}w_{2} + c_{3}w_{1}w_{2} + c_{4}w_{1}^{2} \\
        &h(w_{1},w_{2},w_{3}) &&= c_{5}w_{3} + c_{6}w_{1}w_{3}^{2} + c_{7}w_{2}w_{3},
    \end{aligned}
\end{equation}
where $c_{i} \in \mathbb{R}$ are the component's coefficients. For arbitrary settings of the coefficients, the function $h$ is not always monotone in $w_{k}$. With a rectifier such as $r(w)=\exp(w)$, the third map component $S_{3}$ is given by
\begin{equation}
    S_{3}(w_{1},w_{2},w_{3}) = \underbrace{c_{0} + c_{1}w_{1} + c_{2}w_{2} + c_{3}w_{1}w_{2} + c_{4}w_{1}^{2}}_{\text{non-monotone part }g(w_{1},w_{2})} + 
    \int_{0}^{w_{3}} \exp (\underbrace{c_{5}\omega + c_{6}w_{1}\omega^{2} + c_{7}w_{2}\omega}_{\text{pre-monotone part }h(w_{1},w_{2},w_{3})}) \textrm{d} \omega,
    \label{eq:integrated_rectifier_example}
\end{equation}
which yields a monotone function with respect to $w_3$.
\end{example}

\subsubsection{Monotonicity via separable maps} \label{subsubsec:seperable_maps}

Another computationally efficient way to guarantee monotonicity is to formulate a map component function $S_{k}$ with an additive and separable dependence on $w_k$ \citep{Spantini2022CouplingFiltering}. That is, we write
\begin{equation}
    S_{k}(\w_{1:k}) = g(\w_{1:k-1}) +f(w_{k}),
    \label{eq:separable_map}
\end{equation}
where $g$ is a 
function which depends on the first $k-1$ arguments $\w_{1:k-1}$, and $f$ is a univariate monotone function which depends on only the last argument $w_{k}$. Monotonicity can be ensured by 
\begin{enumerate}
    \item using a linear parameterization for $f$ with only strictly monotone basis functions (e.g., $w_{k}$ and $w_{k}^{3}$), and
    \item constraining the coefficients of the monotone part $f$ to be positive during the map optimization.
\end{enumerate}
\begin{example} 
An example of a separable nonlinear map can be illustrated with a variant of the previous three-variable example, which also uses polynomial basis functions. Let $S_3$ be the map component
\begin{equation}
    S_{3}(w_{1},w_{2},w_{3}) = \underbrace{c_{0} + c_{1}w_{1} + c_{2}w_{2} + c_{3}w_{1}w_{2} + c_{4}w_{1}^{2}}_{\text{non-monotone part }g(w_{1},w_{2})} + \underbrace{c_{5}w_{3} + c_{6}w_{3}^{3}}_{\text{monotone part }f(w_{3})},
\end{equation}
where $c_{i} \in \R$ are the map component's coefficients, and $w_{3}$ and $w_{3}^{3}$ are univariate monotone functions. If $c_5 \geq 0$ and $c_6 \geq 0$, we have $\partial_{w_3} S_3 = f'(w_3) > 0$ and thus $S_3$ is monotone in $w_3$ for all values of the variables $(w_1,w_2) \in \R^2$.
\end{example}

This separable formulation permits efficient map optimization (see Appendix~\ref{sec:AppendixA}), but limits the complexity of the transport map. In particular, since separable map components $S_k$ do not include cross-terms depending on products of $w_{k}$ and $\w_{<k}$, they have more difficulty representing conditionals whose structure (e.g., number of modes) changes as a function of the conditioning variables $\w_{<k}$.

\subsection{Identification of transport maps}\label{subsec:identification}

We next discuss how to find the transport map $\SKR$ that pushes forward a $K$-dimensional target random variable $\w \sim p$ to a $K$-dimensional reference random variable $\z \sim \eta$. In the following, we will assume the target distribution is known only through a collection of i.i.d.\thinspace samples $\{\w^i\}_{i=1}^{N} \sim p$. For example, these samples may arise from the output of a previous operation in the smoothing recursion; see Section~\ref{sec:ensemble_transport_smoothing}. Each target sample $\w^{i}$ can be viewed as one column of a $K$ by $N$ dimensional sample matrix $\W\in\mathbb{R}^{K \times N}$.

We seek the transport map
that minimizes the Kullback--Leibler divergence $D_{\textrm{KL}}(p||\SKR^{\sharp}\eta)$ between the target density $p$ and its approximation, the pullback density $\SKR^{\sharp}\eta$ \citep{Marzouk2017SamplingIntroduction}:
\begin{equation}
\mathcal{D}_{\textrm{KL}}(p||\SKR^{\sharp}\eta) = \int p(\w)\log\frac{p(\w)}{\SKR^{\sharp}\eta(\w)}\textrm{d}\w.
\label{eq:KLD}
\end{equation}
The integration over the target density in this expression can be replaced by a discrete Monte Carlo approximation constructed from individual samples $\w^{i}$. Substituting Equations~\ref{eq:change_of_variables} and \ref{eq:decomposed_S} for the pullback density and using a standard Gaussian reference allows us to derive the following objective function in terms of the map $\SKR$: 
\begin{equation}
    \widehat{\mathcal{J}}(\SKR) = \frac{1}{N}\sum_{i=1}^{N}\sum_{k=1}^{K} \left (\frac{1}{2}S_{k}(\w^{i})^{2} - \log\frac{\partial_{k}S_{k}(\w^{i})}{\partial w_{k}} \right ).
\label{eq:objective_function_full}
\end{equation}
The full derivation of Equation~\ref{eq:objective_function_full} and Equation~\ref{eq:objective_function} is provided in~\citet[Appendix A]{Ramgraber2022underUpdates}. Upon reversing the summations it becomes evident that each summand over $k$ depends only on $S_{k}$, and not on any other map components $S_{\neq k}$. We can consequently solve the full optimization problem by independently minimizing $K$ separate objective functions $\widehat{\mathcal{J}}_{k}(S_{k})$ for $k=1,\ldots,K$, one for each $S_k$:
\begin{equation}
\widehat{\mathcal{J}}_{k}(S_{k}) = \frac{1}{N}\sum_{i=1}^{N} \left (\frac{1}{2}S_{k}(\w^{i})^{2} - \log\frac{\partial_{k}S_{k}(\w^{i})}{\partial w_{k}} \right ).
\label{eq:objective_function}
\end{equation}
We note that this expression also corresponds to a maximum likelihood estimate of the target samples $\W$ over the pullback pdf $\SKR^{\sharp}\eta$, the map's approximation to the target pdf $p$.

The map objective function in Equation~\ref{eq:objective_function} has an intuitive interpretation. Minimizing $\widehat{\mathcal{J}}_{k}$ with respect to the map component $S_k$ attempts, on the one hand, to minimize the first term in the summation by mapping the samples $\w^{i}\sim p$ to values that are close to zero, i.e., to the mode of the reference distribution. On the other hand, the second term in the objective is minimized by maximizing evaluations of the map's derivative, which increases the spread of the samples. The optimal compromise between these two antagonistic elements of the objective depends on the target samples and on constraints imposed by the parameterization adopted for the map. 

In practice, the optimization of Equation~\ref{eq:objective_function} with respect to the parameters of $\SKR$ can be solved efficiently using a quasi-Newton solver~\citep{Baptista2020OnMaps}. The specific optimization objectives for monotone maps that use separable and integrated parameterizations (see Section~\ref{subsec:monotonicity}) are provided in Appendix~\ref{sec:AppendixA} and Appendix~\ref{sec:AppendixB}, 
respectively.

\subsection{Bayesian inference and composite maps}\label{subsec:composite_maps}

The conditioning operation described in Section~\ref{subsec:inv_and_cond} can be used to derive transport versions of Bayesian inference algorithms, including recursive smoothing. This is facilitated if we divide the target variable $\w$ and the related transport map $\SKR(\w)$ into blocks that distinguish time-dependent state and observations variables.

As an example, consider two random vectors of observations $\y_{t}\in\mathbb{R}^{M}$ and states $\x_{t}\in\mathbb{R}^{D}$ at time $t$ with the joint density $p(\x_{t},\y_{t})$. If we wish to condition the states on a particular measurement $\y^{*}_{t}$ of $\y_{t}$, we define the random vector $\w\in\mathbb{R}^{K}$, where $K=M+D$, to be $\w=[\y_{t},\x_{t}]^\top=[y_{t,1},\dots,y_{t,M},x_{t,1},\dots,x_{t,D}]^\top$, where $\y_{t}$ and $\x_{t}$ are viewed as two blocks of $M$ and $D$ scalar variables, respectively. The ordering of observation and state variables in $\w$ enables us to distinguish two corresponding blocks in a transport map $\SKR(\w) = \SKR(\y_{t},\x_{t})$ that relates the target vector $\w$ to a reference vector $\z$:
\begin{equation}
\SKR(\y_{t},\x_{t})=\left[\begin{array}{lr}
    S_{1}(y_{t,1}) \\
    \vdots \\
    S_{M}(y_{t,1},\dots,y_{t,M}) \\[3pt] 
    \hline
    S_{M+1}(y_{t,1},\dots,y_{t,M},x_{t,1}) \\
    \vdots \\
    S_{M+D}(y_{t,1},\dots,y_{t,M},x_{t,1},\dots,x_{t,n})
\end{array}\right]
= \left[\begin{array}{lr}
    \SKR_{\y_{t}}(\y_{t}) \\[3pt] 
    \hline
    \SKR_{\x_{t}}(\y_{t},\x_{t})
\end{array}\right] = \left[\begin{array}{lr}
    \z_{1} \\[3pt] 
    \hline
    \z_{2}
\end{array}\right] = \z.
\label{eq:decomposed_S_block_form}
\end{equation}
The horizontal line in Equation~\ref{eq:decomposed_S_block_form} is a visual aid to separate the two map blocks, and $\z_{1}\sim\eta_{1}\in\mathbb{R}^{M}$ and $\z_{2}\sim\eta_{2}\in\mathbb{R}^{D}$ are blocks of $\z=[\z_{1},\z_{2}]\sim\eta$. This blocked version of the triangular map
has the same form as the map in Equation~\ref{eq:decomposed_S} and may be inverted to sample the joint density $p(\x_{t},\y_{t})$ and its conditionals, including the posterior $p(\x_{t}|\y^{*}_{t})=p(\x_{t},\y_{t}^*)/p(\y^{*}_{t})$, as described in Equations~\ref{eq:marginal_conditional} and~\ref{eq:marginal_conditional_telescopic}. In Section 4.2 we discuss how extensions of this blocked approach can be used to generate samples from the 
posterior density $p(\x_{1:t}|\y_{1:t}^{*})$ that form the basis for the recursive smoothing update introduced in Section~\ref{sec:introduction}.

There are two ways to use the block transport map to sample from a target  posterior density $p(\x_{t}|\y_{t}^{*})$ of states $\x_{t}$ conditioned on specific observations $\y_{t}^{*}$ at time $t$:

\begin{enumerate}
    \item \textbf{Pullback} (reference-to-conditional): Draw fresh independent reference samples $\z_{2}^{i} \sim \eta_{2}$ from the standard Gaussian reference distribution. Then use the observation-conditioned inverse map $\SKR_{\x_{t}}^{-1}(\z_{2}^i;\y_{t}^{*})$ to transform  $\z_{2}^{i}$ into samples from the target, or
    \item \textbf{Composite} (joint-to-conditional): Use the forward map $\widetilde{\z}_{2}^{i} = \SKR_{\x_{t}}(\y_{t}^{i},\x_{t}^{i})$ to transform (training) samples $(\y_{t}^{i},\x_{t}^{i})$ from the joint distribution $p(\y_{t},\x_t)$ into samples $\widetilde{\z}_{2}^{i}$ from the pushforward $\SKR_{\#}p$. We then evaluate the inverse map  $\SKR_{\x_{t}}^{-1}(\widetilde{\z}_{2};\y_{t}^{*})$ at these approximate reference $\widetilde{\z}_{2}^{i}$ samples given a specific observation $\y_{t}^{*}$.
\end{enumerate} 

The latter strategy based on composite maps bears some notable advantages over the former since it works well with simpler (e.g., lower order) nonlinear maps. This is because composite maps may partially preserve features of the target distribution $p$ not captured by the pullback density of imperfect lower-complexity maps $\SKR$~\citep{Spantini2022CouplingFiltering}.
Furthermore, the composite approach does not require sampling extraneous reference samples and is especially convenient in recursive Bayesian inference applications, where we wish to convert the prior ensemble into a posterior ensemble.

The composite approach can be concisely described with a composite map $\T_{\y_{t}^{*}}$ defined as \citep{Spantini2022CouplingFiltering}:
\begin{equation}
    \x_{t}^{*} = 
    \T_{\y_{t}^{*}}(\y_{t},\x_{t}) \coloneqq 
    \SKR_{\x_{t}}^{-1}(\y_{t}^{*},\cdot)\circ\SKR_{\x_{t}}(\y_{t},\x_{t}),
    \label{eq:composite_map}
\end{equation}
where $\x_{t}^{*}$ is distributed according to the conditional density $p(\x_{t}|\y_{t}^{*})$. The superscript $*$ on $\x_{t}^{*}$ is used to emphasize that the random variable produced by the composite map depends on a particular measurement value $\y_{t}^{*}$. Since the composite map extracts reference samples only through the lower map block $\SKR_{\x_{t}}$ and conditionally inverts these samples with $\SKR^{-1}_{\x_{t}}$ again, it does not use the upper map block $\SKR_{\y_{t}}$. So long as our interest lies in extracting conditional distributions, we do not need to build, optimize, or even define $\SKR_{\y_{t}}$. This is consistent with the conditioning approach illustrated in Figure~\ref{fig:conditioning}B, where the first inversion step from Figure~\ref{fig:conditioning}A is skipped. 

In the example above, we consider an inference problem where all members of the state ensemble $\X_t$ are conditioned on the same realized value of the observation $\y_t^*$. We note that it is also possible to condition each ensemble member $\x_{t}^{i}$ on a separate value of the observation (or any other conditioning variable taking the place of $\y_t$). For example, as we shall see in Section~\ref{subsec:backward_EnTS}, in backward smoothing we condition $\X_t$ on $\X_{t+1}^{*}$, an ensemble of individual realizations of the updated state vector $\x^{*,i}_{t+1}$. The $i$th member of $\X_t$ is thus conditioned on the value of the $i$th member of $\X_{t+1}^{*}$.

\section{Ensemble transport smoothers (EnTS)} \label{sec:ensemble_transport_smoothing}
We now discuss how the transport map approach to conditional sampling in Section~\ref{sec:transport_maps} can be applied specifically to solve smoothing problems. The general goal of Bayesian smoothers is to characterize the posterior density $p(\x_{1:t}|\y_{1:t}^{*})$ of a sequence of time-dependent state vectors $\x_{1:t}$ given a concurrent sequence of time-dependent measurements $\y_{1:t}^{*}$. In principle, this high-dimensional inference problem could be solved with transport methods as a single (batch) conditioning problem in which all the states $\x_{1:t}$ are simultaneously updated given all of the measurements $\y_{1:t}^{*}$. However, batch updates can require manipulating large transport maps when the smoothing window is long.

It is often both easier and more computationally efficient to carry out
the smoothing procedure recursively, in a series of forecast and inference steps. 
Recursive algorithms incrementally condition joint distributions that have been informed by earlier data, such as $p(\y_{t},\x_{1:t}|\y_{1:t-1}^{*})$. A recursive approach improves stability by relying on more accurate conditional state forecasts and it provides a natural way to exploit conditional independence properties that yield sparse transport maps. These are the primary reasons why we consider only recursive ensemble smoothers in the rest of our discussion.

\begin{figure}[!ht]
  \centering
  \includegraphics[width=\textwidth]{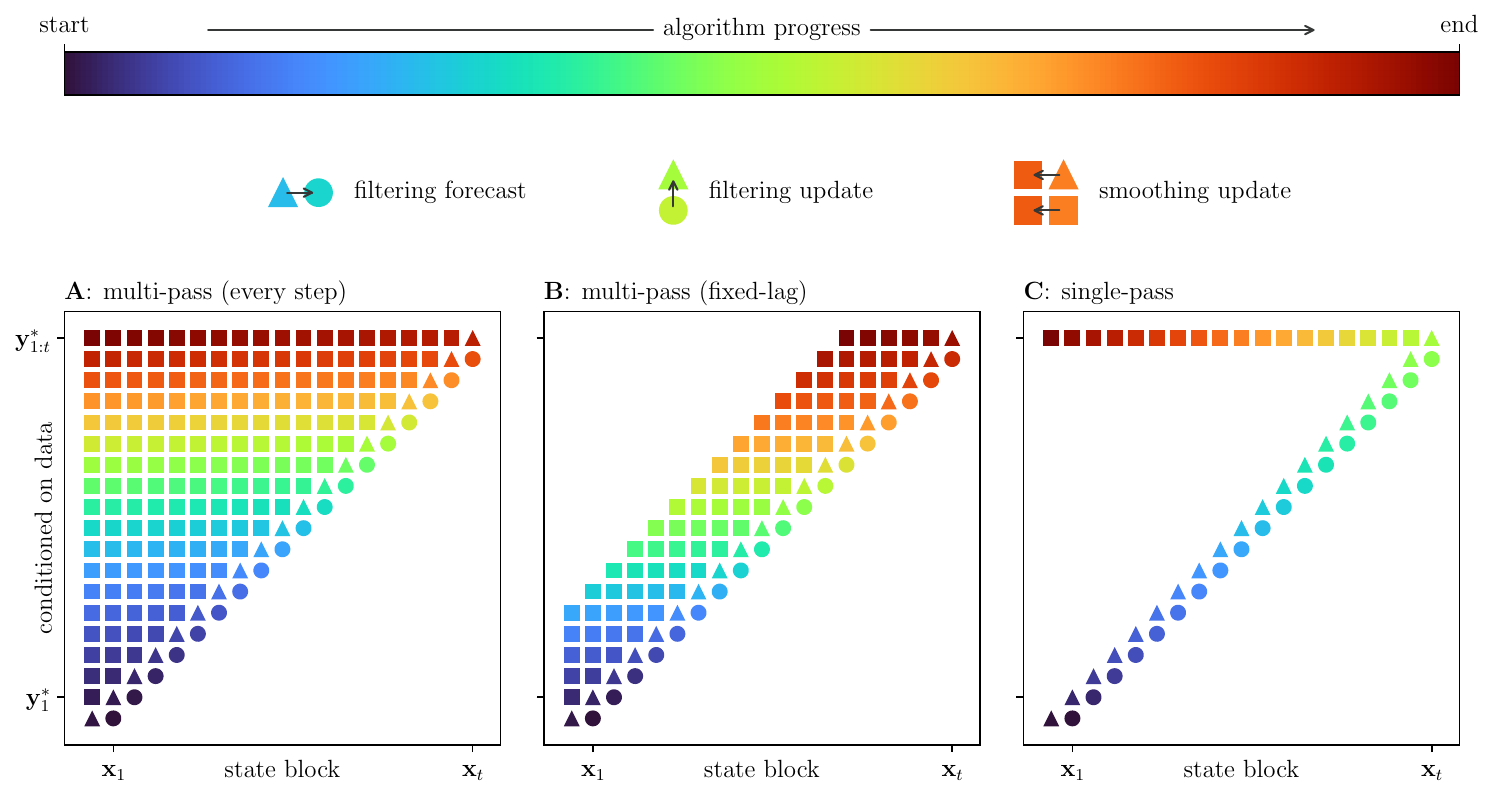}
  \caption{Different backward smoothers. Each smoother is based on a forward filtering pass along the diagonal and first sub-diagonal, representing a filtering update and a forecast, respectively. The classic expanding window multi-pass smoother (A) performs a backward smoothing update pass through all previous times whenever a new observation becomes available. The fixed-lag smoother (B) performs smoothing updates only within a specified lag window that ends at the current time. The fixed window single-pass smoother (C) uses a single backward smoothing update pass that starts at the end of the forward filtering window and moves through all preceding times.}
  \label{fig:backward_smoother_types}
\end{figure}

\citet{Ramgraber2022underUpdates} identifies and compares several different smoothing strategies, including \textit{dense smoothers} (applying the update with no regards to conditional independence), \textit{forward smoothers} (exploiting conditional independence in a serial update forward along the graph), and \textit{backward smoothers} (exploiting conditional independence in a serial update backward along the graph). When the composite map in Equation~\ref{eq:composite_map} that performs the inference step is constrained to be linear, the transport versions of the first and last algorithms correspond to methods like the Ensemble Kalman Smoother (EnKS) and the Ensemble Rauch-Tung-Striebel Smoother (EnRTSS). The companion paper demonstrates that backward ensemble transport formulations are generally better-suited for nonlinear smoothing than dense and forward smoothers, for the following reasons:
\begin{itemize}
    \item \textbf{Robustness}: The updates of backward smoothers are more robust to spurious correlations, and thus yield better tracking performance for small ensemble sizes.
    \item \textbf{Adaptability}: 
    A backwards smoother can handle selective multi-pass updates such as shown in Figure~\ref{fig:backward_smoother_types}B more efficiently than the joint analysis counterpart.
\end{itemize}

Backwards smoothers can be further divided into \textit{multi-pass} and \textit{single-pass} smoothers, which are appropriate in different applications. Multi-pass smoothers update all states in multiple backward passes that start at specified measurement times, as shown in Figures~\ref{fig:backward_smoother_types}A and \ref{fig:backward_smoother_types}B. Note that the color bar in this figure indicates that the recursion starts at the beginning of the time window (from dark blue) and moves forward (to dark red) with a filtering pass that updates the current state with a new measurement at each time step. Previous states are updated in backward smoothing passes that start at every time step in Option A and at selected time steps in Option B. This multi-pass approach is appropriate when there is an \textit{expanding time window}. In this case, which might be encountered in a real-time application, measurements are continually added at the end of the smoothing window $1:t$ so the final time $t$ is always increasing. 

The computational effort required by an expanding window smoother continually grows over time. An approximation that addresses this dilemma is a \textit{fixed-lag multi-pass} smoother (Figure~\ref{fig:backward_smoother_types}C), which only updates the most recent states. This simplification is justified in many situations by the gradual decay in the information that a new measurement $\y_{t}$ provides about a previous state $\x_{s}$, as the time difference between $t$ and $s$ increases. 

A \textit{single-pass} smoother (Figure~\ref{fig:backward_smoother_types}D) requires much less computational effort than multi-pass alternatives since it updates all states in one backwards sweep carried out at the end of a fixed length smoothing window. \citet{Ramgraber2022underUpdates} investigates all of these backward smoothing algorithms for the special case of linear updates. In the following sections we present transport-based formulations of the multi-pass and single-pass backwards smoothers illustrated in Figure~\ref{fig:backward_smoother_types} for non-Gaussian problems with nonlinear updates.

\begin{figure}[!ht]
  \centering
  \includegraphics[width=\textwidth]{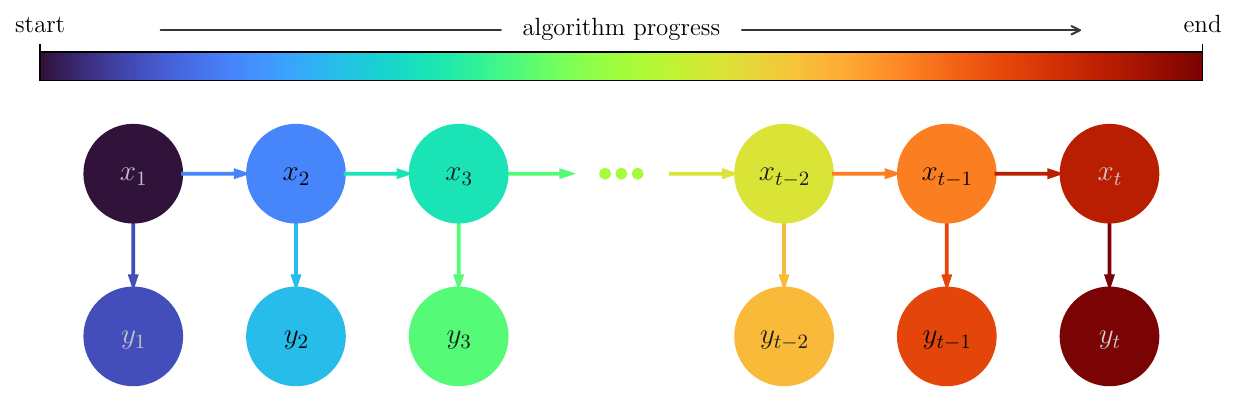}
  \caption{Markov structure of a state space model. The joint density of states and observations $p(\x_{1:t},\y_{1:t})$ is defined as the product of the initial prior $p(\x_1)$, the transition density $p(\x_s|\x_{s-1})$ and the likelihood function $p(\y_s|\x_s)$ for $s = 1,\dots,t$.} 
  \label{fig:Markovian_graph}
\end{figure}

Before discussing map implementations and sample generation for these smoothers, we consider decompositions (or factorizations) of the joint forecast density $p(\y_{t},\x_{1:t}|\y_{1:t-1}^{*})$ that indicate how the required maps should be structured. These decompositions incorporate the conditional independence relationships between states and measurements that are described by the hidden Markov Model (HMM) in Figure~\ref{fig:Markovian_graph} \citep{ihler2007graphical,elliott2008hidden}. This graphical model implies the following conditional independencies:
\begin{itemize}
    \item \textbf{Markovian states}: $\x_{1:s-1}\perp \!\!\! \perp \x_{s+1:t} \; | \; \x_{s} \; \forall s = 2,\dots,t-1$. Given its immediate neighbour $\x_{s-1}$ or $\x_{s+1}$, each state $\x_{s}, \; s = 2,\dots,t-1,$ is conditionally independent of states beyond this neighbour, i.e., $\x_{1:s-2}$ or $\x_{s+2:t}$.
    \item \textbf{Conditionally independent observations}: $\y_{s}\perp \!\!\! \perp \x_{j\neq s} \; | \; \x_{s} \; \forall j,s = 1,\dots,t$. Given the corresponding hidden state $\x_{s}$, each observation $\y_{s}$ is conditionally independent of all other states $\x_{j\neq s}$.
\end{itemize}

We use these conditional independence relationships to express the forecast density for the states and latest observation as a product of the previous smoothing density $p(\x_{1:t-1}|\y_{1:t-1}^{*})$, the transition density $p(\x_{t}|\x_{t-1})$, and the likelihood function $p(\y_{t}|\x_{t})$, as follows:
\begin{align}
    p(\y_{t},\x_{1:t}|\y_{1:t-1}^{*}) &= p(\x_{1:t-1}|\y_{1:t-1}^{*})p(\x_{t}|\x_{1:t-1},\y_{1:t-1}^{*})p(\y_{t}|\x_{1:t},\y_{1:t-1}^{*}) \nonumber \\
    &= p(\x_{1:t-1}|\y_{1:t-1}^{*})p(\x_{t}|\x_{t-1})p(\y_{t}|\x_{1:t},\y_{1:t-1}^{*}) && \left[\x_{t}\perp \!\!\! \perp \x_{1:t-2},\y_{1:t-1} \; | \; \x_{t-1}\right] \nonumber \\
    &= p(\x_{1:t-1}|\y_{1:t-1}^{*})p(\x_{t}|\x_{t-1})p(\y_{t}|\x_{t}) && \left[\y_{t}\perp \!\!\! \perp \y_{1:t-1},\x_{1:t-1} \; | \; \x_{t}\right].
\label{eq:forecast_augmented_smoothing_dist}
\end{align}
The factorization in Equation~\ref{eq:forecast_augmented_smoothing_dist} provides a way to generate samples needed to learn the map given a previous smoothing ensemble. 
We can also factorize this forecast density to recover a general expression for the new smoothing density $p(\x_{1:t}|\y_{t},\y_{1:t-1}^{*})$, as indicated by the second equality of the following expression:
\begin{equation}
    p(\x_{1:t-1}|\y_{1:t-1}^{*})p(\x_{t}|\x_{t-1})p(\y_{t}|\x_{t}) = p(\y_{t},\x_{1:t}|\y_{1:t-1}^{*}) = p(\y_{t}|\y_{1:t-1}^{*})p(\x_{1:t}|\y_{t},\y_{1:t-1}^{*}).
    \label{eq:Bayes_factorized_joint}
\end{equation}
Following the derivation of Bayes' theorem, we divide both sides by $p(\y_{t}|\y_{1:t-1}^{*})$ to obtain a recursive expression for the smoothing density:
\begin{equation}
    p(\x_{1:t}|\y_{t},\y_{1:t-1}^{*})= \frac{p(\y_{t},\x_{1:t}|\y_{1:t-1}^{*})}{p(\y_{t}|\y_{1:t-1}^{*})} .
    \label{eq:Bayes_factorized}
\end{equation}

Ensemble transport methods can sample this smoothing distribution by locating the observation block $\y_{t}$ of the forecast map in the uppermost entries of the map, then using a composite map approach to sample the conditional (see Section~\ref{subsec:inv_and_cond}). Different orderings of the remaining state blocks $\x_{1:t}$ lead to different factorizations of the forecast density (Equation~\ref{eq:forecast_augmented_smoothing_dist}) that, in turn, yield different smoothing algorithms. In particular, a backward-in-time ordering $\w=(\y_{t},\x_{t},\x_{t-1},\dots,\x_{1})$ yields a generic backward smoother, which can be further sparsified using the HMM's conditional independence properties:
\begin{equation}
    p(\x_{t:1}|\y_{t},\y_{1:t-1}^{*}) = p(\x_{t}|\y_{t},\y_{1:t-1}^{*})p(\x_{t-1}|\y_{t},\x_{t},\y_{1:t-1}^{*})\cdots p(\x_{1}|\y_{t},\x_{2:t},\y_{1:t-1}^{*})
    \label{eq:posterior_factorized}
\end{equation}
This posterior is sampled by the inverse of the lower block $\SKR_{\x_{t:1}}$ of the full map $\SKR=[\SKR_{\y_{t}},\SKR_{\x_{t:1}}]=[\SKR_{\y_{t}},\SKR_{\x_{t}},\dots,\SKR_{\x_{1}}]$ for the joint forecast density $p(\y_{t},\x_{1:t}|\y_{1:t-1}^{*})$, as described in Equation~\ref{eq:marginal_conditional}. Using the HMM's conditional independence properties (detailed further below), Equation~\ref{eq:posterior_factorized} can be sparsified to:
\begin{equation}
    p(\x_{t:1}|\y_{t},\y_{1:t-1}^{*}) = p(\x_{t}|\y_{t},\y_{1:t-1}^{*})p(\x_{t-1}|\x_{t},\y_{1:t-1}^{*})\cdots p(\x_{1}|\x_{2},\y_{1}^{*})
    \label{eq:posterior_supersparse}
\end{equation}
Below we derive three different backward smoothing algorithms which, beginning with the factorization in Equation~\ref{eq:posterior_factorized}, exploit progressively more conditional independence until we obtain a backward recursion that fully reflects the factorization in Equation~\ref{eq:posterior_supersparse}. 

\subsection{Expanding-window multi-pass backward EnTS}\label{subsec:backward_EnTS}

Using the Markov properties stated above, we can drop unnecessary dependencies on $\y_t$ and on non-neighbouring future states from the appropriate terms in Equation~\ref{eq:posterior_factorized}. The resulting sparse decomposition of the joint density is used to derive the \textit{expanding window multi-pass} form of the backwards EnTS:
\begin{equation}
    p(\x_{t:1}|\y_{t},\y_{1:t-1}^{*}) = 
    \underbrace{p(\x_{t}|\y_{t},\y_{1:t-1}^{*})}_{\text{filtering}}
    \underbrace{p(\x_{t-1}|\x_{t},\y_{1:t-1}^{*})}_{\text{backward smoothing}}\cdots 
    \underbrace{p(\x_{1}|\x_{2},\y_{1:t-1}^{*})}_{\text{backward smoothing}}.
    \label{eq:backwards_graph_multipass}
\end{equation}
This factorization is sampled by the inverse of the sparse lower block $\SKR_{\x_{t:1}}$ of the triangular forecast map for the joint density in Equation~\ref{eq:forecast_augmented_smoothing_dist}:
\begin{equation}
\SKR(\y_{t},\x_{t:1})=\left[\begin{array}{lr}
    \SKR_{\y_{t}}(\y_{t}) \\
    \hline 
    \SKR_{\x_{t:1}}(\y_{t},\x_{t:1})
\end{array}\right]=\left[\begin{array}{lr}
    \SKR_{\y_{t}} (\y_{t}) \\
    \hline 
    \SKR_{\x_{t}} (\y_{t},\x_{t}) \\
    \SKR_{\x_{t-1}} (\x_{t},\x_{t-1}) \\
    \vdots \\
    \SKR_{\x_{2}} (\x_{3},\x_{2}) \\
    \SKR_{\x_{1}} (\x_{2},\x_{1})
\end{array}\right]\begin{array}{lr}
    \\
    \big\}\text{ filtering} \\
    \big\}\text{ backward smoothing} \\
    \vdots \\
    \big\}\text{ backward smoothing} \\
    \big\}\text{ backward smoothing} \\
\end{array}.
\label{eq:decomposed_S_EnTS}
\end{equation}
We can sample the recursive posterior density with this map using the composite map approach summarized  in Equation~\ref{eq:composite_map}. 
The association of the terms and map component blocks in Equations~\ref{eq:backwards_graph_multipass} and \ref{eq:decomposed_S_EnTS} with separate operations reflects the fact that these sparse blocks (and their corresponding terms in Equation~\ref{eq:backwards_graph_multipass}) also have alternative interpretations. Instead of parsing them as components of a single, large map, they can also be interpreted as components of a sequence of multiple, smaller, overlapping maps: 
\begin{equation}
\begin{aligned}
&\text{at time }t: &&\SKR(\y_{t},\x_{t})&&=\begin{bmatrix*}[l]
    \SKR_{\y_{t}} (\y_{t}) \\
    \SKR_{\x_{t}} (\y_{t},\x_{t})
\end{bmatrix*}\begin{array}{lr}
    \\
    \big\}\text{ filtering} \\
\end{array}
\\
&\text{at times }s<t: &&\SKR(\x_{s+1},\x_{s})&&=\begin{bmatrix*}[l]
    \SKR_{\x_{s+1}} (\x_{s+1}) \\
    \SKR_{\x_{s}} \;\;\; (\x_{s+1},\x_{s})
\end{bmatrix*}\begin{array}{lr}
    \\
    \big\}\text{ backward smoothing} 
\end{array},
\end{aligned}
\label{eq:backward_smoother_subdivided}
\end{equation}
where the inverses of the lower map component blocks $\SKR_{\x_{t}}$ and $\SKR_{\x_{s}}$ realize the filtering and backward smoothing updates, respectively. In this sequence of operations, the conditional inversion (see Section~\ref{subsec:inv_and_cond}) of $\SKR_{\x_{t}}$ on $\y_{t}^{*}$ provides the filtering marginal samples $\x_{t}^{*}\sim p(\x_{t}|\y_{1:t}^{*})$, which in turn serve as conditioning input for the inversion of $\SKR_{\x_{t-1}}$, yielding $\x_{t-1}^{*}$. This serves as input for the next backward smoothing operation, and so on, until the first state block $\x_{1}$ is updated. The perspectives of Equation~\ref{eq:decomposed_S_EnTS} and Equation~\ref{eq:backward_smoother_subdivided} give equivalent results. The pseudo-code provided in Algorithm~\ref{alg:backward_smoother_multi_pass} adopts the formulation of Equation~\ref{eq:backward_smoother_subdivided}.

\begin{algorithm}[!ht]
\SetAlgoLined
\DontPrintSemicolon

 \textbf{Input}: $N$ prior samples $\X_1\sim p(\x_{1})$, the forecast model $p(\x_{s}|\x_{s-1})$, the observation model $p(\y_{s}|\x_{s})$, and a stream of observations $\y_{1:t}^{*}$.
 
 \For{$s = 1:t$}{
   
    \If{$ s \neq 1$} {
    
        \textit{Forecast state}: Sample $\x_{s}^{i} \sim p(\x_{s}|\x_{s-1}^{*i}), \  \forall \, i=1,\dots,N$\;
    }
    
    \textit{Forecast observation}: Sample $\y_{s}^{i} \sim p(\y_{s}|\x_{s}^{i}), \  \forall \, i=1,\dots,N$\;
    
    \textit{Filtering step}:\;
    a) Build and optimize the map component block $\SKR_{\x_s}(\y_{s},\x_{s})$ using $(\Y_s,\X_s)$\;
    b) Push forward $\Z_{s}=\SKR_{\x_s}(\Y_{s},\X_{s})$\;
    c) Pull back $\X_{s}^{*}=\SKR_{\x_s}^{-1}(\y_{s}^{*};\Z_{s})$\;

    \textit{Prepare map samples}:\;
    Define $\X_{1:s-1}^{\text{prev}} \coloneqq \X_{1:s-1}^{*}$ (previous smoothing marginals) \; 
    Define $\X_{2:s}^{\text{next}} \coloneqq [\X_{2:s-1}^{*},\X_{s}]^{\top}$ ($\X_{2:s-1}^{*}$: previous smoothing marginals, $\X_{s}$: filtering forecast)\; 

    \For{$r = (s-1):1$}{
        \textit{Backward smoothing step}:\;
        a) Build and optimize the map component block $\SKR_{\x_r}(\x_{r+1}^{\text{next}},\x_{r}^{\text{prev}})$ using $(\X_{r+1}^{\text{next}},\X_r^{\text{prev}})$\;
        b) Push forward $\Z_{r}=\SKR_{\x_r}(\X_{r+1}^{\text{next}},\X_{r}^{\text{prev}})$\;
        c) Pull back $\X_{r}^{*}=\SKR_{\x_r}^{-1}(\X_{r+1}^{*};\Z_{r})$\;
     }
    \textit{Update samples}: Set $\X_{1:s} = \X_{1:s}^*$\;
    
    }
    \caption{Multi-pass backward EnTS}
    \label{alg:backward_smoother_multi_pass}
    
\end{algorithm}

\subsection{Fixed-lag multi-pass backward EnTS} 

It is also possible to derive a \textit{fixed-lag multi-pass} backward EnTS from the previous method by omitting all updates, at time $t$, to state blocks more than $L$ steps in the past. This gives the following decomposition of the joint distribution:
\begin{equation}
    p(\x_{t:1}|\y_{1:t-1}^{*}) \approx 
    \underbrace{p(\x_{t}|\y_{t},\y_{1:t-1}^{*})}_{\text{filtering}}
    \underbrace{p(\x_{t-1}|\x_{t},\y_{1:t-1}^{*})}_{\text{backward smoothing}}\cdots 
    \underbrace{p(\x_{t-L}|\x_{t-L+1},\y_{1:t-1}^{*})}_{\text{backward smoothing}}
    \underbrace{p(\x_{t-L-1:1}|\y_{1:t-1}^*)}_{\text{inherited }}.
    \label{eq:backwards_graph_multipass_fixed_lag_app}
\end{equation}
Here again, the update operations are indicated  with underbraces. The last right-hand side term in Equation~\ref{eq:backwards_graph_multipass_fixed_lag_app} contains states beyond the smoothing lag $L$, which are  unaltered in this inference step and thus inherited from previous smoothing passes. Separated into distinct operations, the fixed-lag smoother can be defined as:
\begin{equation}
\begin{aligned}
&\text{at time }t: &&\SKR(\y_{t},\x_{t})&&=\begin{bmatrix*}[l]
    \SKR_{\y_{t}} (\y_{t}) \\
    \SKR_{\x_{t}} (\y_{t},\x_{t})
\end{bmatrix*}&&\begin{array}{lr}
    \\
    \!\!\!\big\}\text{ filtering} \\
\end{array} \\
&\text{at times }t-L\geq s<t: &&\SKR(\x_{s+1},\x_{s})&&=\begin{bmatrix*}[l]
    \SKR_{\x_{s+1}} (\x_{s+1}) \\
    \SKR_{\x_{s}} \;\;\; (\x_{s+1},\x_{s})
\end{bmatrix*}&&\begin{array}{lr}
    \\
    \!\!\!\big\}\text{ backward smoothing} 
\end{array} \\
&\text{at times }s<t-L: &&\z_{s} &&=\x_{s} &&\big\}\text{ inherited (not updated)} .
\end{aligned}
\label{eq:backward_smoother_subdivided_fixed_lag}
\end{equation}
This truncation of the update is often justified pragmatically, arguing that beyond time $s < t-L$, we have $\x_{s}\approx\x_{s}^{*}$. In consequence, the update has virtually no effect for earlier state blocks, and can omitted at no significant loss in fidelity. This is explored in more detail in \citet{Ramgraber2022underUpdates}. Pseudo-code for this smoothing variant is provided in Algorithm~\ref{alg:backward_smoother_multi_pass_fixed_lag}.

\begin{algorithm}[!ht]
\SetAlgoLined
\DontPrintSemicolon

 \textbf{Input}: $N$ prior samples $\X_1\sim p(\x_{1})$, the forecast model $p(\x_{s}|\x_{s-1})$, the observation model $p(\y_{s}|\x_{s})$,  a stream of observations $\y_{1:t}^{*}$, and \textcolor{blue_custom}{an optional lag parameter $L \in \mathbb{N}^{+}$, else $L = t$}.
 
 \For{$s = 1:t$}{
   
    \If{$ s \neq 1$} {
    
        \textit{Forecast state}: Sample $\x_{s}^{i} \sim p(\x_{s}|\x_{s-1}^{*i}), \  \forall \, i=1,\dots,N$\;
    }
    
    \textit{Forecast observation}: Sample $\y_{s}^{i} \sim p(\y_{s}|\x_{s}^{i}), \  \forall \, i=1,\dots,N$\;
    
    \textit{Filtering step}:\;
    a) Build and optimize the map component block $\SKR_{\x_s}(\y_{s},\x_{s})$ using $(\Y_s,\X_s)$\;
    b) Push forward $\Z_{s}=\SKR_{\x_s}(\Y_{s},\X_{s})$\;
    c) Pull back $\X_{s}^{*}=\SKR_{\x_s}^{-1}(\y_{s}^{*};\Z_{s})$\;

    \textit{Prepare map samples}:\;
    Define $\X_{1:s-1}^{\text{prev}} \coloneqq \X_{1:s-1}^{*}$ (previous smoothing marginals) \; 
    Define $\X_{2:s}^{\text{next}} \coloneqq [\X_{2:s-1}^{*},\X_{s}]^{\top}$ ($\X_{2:s-1}^{*}$: previous smoothing marginals, $\X_{s}$: filtering forecast)\; 

    \For{$r = (s-1):\textcolor{blue_custom}{\max (1, s-L)}$}{
        \textit{Backward smoothing step}:\;
        a) Build and optimize the map component block $\SKR_{\x_r}(\x_{r+1}^{\text{next}},\x_{r}^{\text{prev}})$ using $(\X_{r+1}^{\text{next}},\X_r^{\text{prev}})$\;
        b) Push forward $\Z_{r}=\SKR_{\x_r}(\X_{r+1}^{\text{next}},\X_{r}^{\text{prev}})$\;
        c) Pull back $\X_{r}^{*}=\SKR_{\x_r}^{-1}(\X_{r+1}^{*};\Z_{r})$\;
     }
    \textit{Update samples}: Set $\X_{1:s} = \X_{1:s}^*$\;
    
    }
    \caption{Fixed-lag multi-pass backward EnTS. Differences to Algorithm~\ref{alg:backward_smoother_multi_pass} are emphasized in \textcolor{blue_custom}{blue}.}
    \label{alg:backward_smoother_multi_pass_fixed_lag}
    
\end{algorithm}

\subsection{Fixed-window single-pass backward EnTS}

We can exploit additional sparsity if we extend the two independence properties above by noting that $\x_{s}\perp \!\!\! \perp \y_{j > s} \; | \; \x_{s+1} \; \forall j,s = 1,\dots,t-1$. That is, given a future state block $\x_{s+1}$, the preceding state block $\x_{s}$ is conditionally independent of all future observation predictions $\y_{j > s}$. This allows us to write each term on the right-hand side of Equation~\ref{eq:backwards_graph_multipass} as $p(\x_s|\x_{s+1},\y_{1:t-1}^*) = p(\x_s|\x_{s+1},\y_{1:s}^*)$, which yields Equation~\ref{eq:posterior_supersparse}, further annotated as follows:
\begin{equation}
    p(\x_{t:1}|\y_{t},\y_{1:t-1}^{*}) = 
    \underbrace{p(\x_{t}|\y_{t},\y_{1:t-1}^{*})}_{\text{filtering}}
    \underbrace{p(\x_{t-1}|\x_{t},\y_{1:t-1}^{*})}_{\text{backward smoothing}} \underbrace{p(\x_{t-2}|\x_{t-1},\y_{1:t-2}^{*})}_{\text{backward smoothing}}\cdots 
    \underbrace{p(\x_{1}|\x_{2},\y_{1}^{*})}_{\text{backward smoothing}}.
    \label{eq:backwards_graph_singlepass}
\end{equation}

Under linear-Gaussian assumptions, this factorization underlies the Rauch-Tung-Striebel smoother \citep{Rauch1965MaximumSystems}. Despite the apparent similarity to Equation~\ref{eq:backwards_graph_multipass}, this expression could only be expressed as a single, batch map operation as in Equation~\ref{eq:decomposed_S_EnTS} if we were to condition a joint distribution $p(\y_{1:t},\x_{1:t})$ on all data points $\y_{1:t}^{*}$ at once. For all the reasons against batch operations listed above, this is sub-optimal. Instead, we can parse Equation~\ref{eq:backwards_graph_singlepass} as a sequence of smaller, independent update operations, as in Equation~\ref{eq:backward_smoother_subdivided}:
\begin{equation}
\begin{aligned}
&\text{at time }t: &&\SKR(\y_{t},\x_{t})&&=\begin{bmatrix*}[l]
    \SKR_{\y_{t}} (\y_{t}) \\
    \SKR_{\x_{t}} (\y_{t},\x_{t})
\end{bmatrix*} && \text{where}\quad(\y_{t},\x_{t}) \sim p(\y_{t},\x_{t}|\y_{1:t-1}^{*})
\\
&\text{at times }s<t: &&\SKR(\x_{s+1},\x_{s})&&=\begin{bmatrix*}[l]
    \SKR_{\x_{s+1}} (\x_{s+1}) \\
    \SKR_{\x_{s}} \;\;\; (\x_{s+1},\x_{s})
\end{bmatrix*} && \text{where}\quad(\x_{s+1},\x_{s}) \sim p(\x_{s+1},\x_{s}|\y_{1:s}^{*}),
\end{aligned}
\label{eq:backward_smoother_subdivided_single_pass}
\end{equation}
where the necessary samples for the backward smoothing joint distributions are obtained during the filtering pass as $p(\x_{s+1},\x_{s}|\y_{1:s}^{*})=p(\x_{s}|\y_{1:s}^{*})p(\x_{s+1}|\x_{s})$, that is to say by concatenating filtering analysis samples from $p(\x_{s}|\y_{1:s}^{*})$ at time $s$ with filtering forecast samples from $p(\x_{s+1}|\x_{s})$ at time $s+1$. This makes it possible to realize the backward smoother with only a single backward pass following a preceding forward filtering pass, rather than requiring the sequence of recursive backwards sweeps used in the multi-pass variant. Pseudo-code for this method is provided in Algorithm~\ref{alg:backward_smoother_single_pass}.

\begin{algorithm}[!ht]
\SetAlgoLined
\DontPrintSemicolon
 \vspace{2 pt}

\textbf{Input}: $N$ prior samples $\X_1\sim p(\x_{1})$, forecast model $p(\x_{s}|\x_{s-1})$, observation model $p(\y_{s}|\x_{s})$, and a stream of observations $\y_{1:t}^{*}$.
 
 \For{$s = 1:t$}{
   
    \If{$ s \neq 1$} {
        \textit{Forecast state}:
        Sample $\x_{s}^{i} \sim p(\x_{s}|\x_{s-1}^{*i}),\  \forall \, i=1,\dots,N$\;
    }
    
    \textit{Forecast observation}:
    Sample $\y_{s}^{i} \sim p(\y_{s}|\x_{s}^{i}),\  \forall \, i=1,\dots,N$\;
    
    \textit{Filtering step}:\;
    a) Build and optimize the map component block $\SKR_{\x_s}(\y_{s},\x_{s})$ using $(\Y_s,\X_s)$\;
    b) Push forward $\Z_{s}=\SKR_{\x_s}(\Y_{s},\X_{s})$\;
    c) Pull back $\X_{s}^{*}=\SKR_{\x_s}^{-1}(\y_{s}^{*};\Z_{s})$\;
    }
    
    \textit{Update samples}: Set $\X_{1:s} = \X_{1:s}^*$\;

    \For{$r = t-1:1$}{
        \textit{Backward smoothing step}:\;
        a) Build and optimize the map component block $\SKR_{\x_r}(\x_{r+1},\x_{r})$ using $(\X_{r+1},\X_r)$\;
        b) Push forward $\Z_{r}=\SKR_{\x_r}(\X_{r+1},\X_{r})$\;
        c) Pull back $\X_{r}^{*}=\SKR_{\x_r}^{-1}(\X_{r+1}^{*};\Z_{r})$\;
    }
 
    \caption{Single-pass backward ensemble transport smoother}
    \label{alg:backward_smoother_single_pass}
\end{algorithm}

\subsection{Computational demand}

The backward smoothers defined in the preceding sections each have different computational cost. A classic multi-pass backward smoother, initiating updates at every time-step, demands a total of $ t(t + 1) / 2$ update operations (including filtering). For longer timeseries, its fixed-lag variant is substantially cheaper, requiring only $tL - L(L-1)/2$ update operations. The most computationally efficient option by far is the single-pass smoother, which demands only $2t-1$ update operations. Due to its computational efficiency, we will use this single-pass formulation (Equation~\ref{eq:backwards_graph_singlepass}) in the experiments discussed in Section 5.

\section{Numerical experiments} \label{sec:experiments}
In this section we demonstrate the performance of the nonlinear single-pass backward EnTS in various scenarios with non-Gaussian distributions: a one-dimensional and bimodal system (Section~\ref{subsec:bimodal_sine}), the three-dimensional Lorenz-63 system (Section~\ref{subsec:L63}), and the 40-dimensional Lorenz-96 system (Section~\ref{subsec:L96}).

The Python code to reproduce the experiments and figures in this study is provided in the GitHub repository: \url{https://github.com/MaxRamgraber/Ensemble-Transport-Smoothing-Part-II}. The triangular transport toolbox we used in this study is available at \url{https://github.com/MaxRamgraber/Triangular-Transport-Toolbox}.

\subsection{Bimodal sine}\label{subsec:bimodal_sine}

As a first illustration of the benefits of nonlinear maps, we consider a linear system with a two-dimensional state $\x(t)=(x^{a}(t),x^{b}(t))$ that generates a sinusoidal output. The state evolves in time according to the coupled ODE
\begin{equation}
    \frac{d x^{a}}{d t} = x^{b}, \quad \frac{d x^{b}}{d t} = -\omega^{2}x^{a},
\end{equation}
where $\omega$ is the sine wave's angular frequency and the initial condition is $\x(0) = (0,\omega)$. We integrate this system analytically over a time interval of length $\Delta t$ to obtain the recursive discrete-time state equations:
\begin{equation}
    \begin{aligned}
    x_{s}^{a} = x^{a}({t_{s-1}+\Delta t}) &= \cos(\omega\Delta t)x_{s-1}^{a}+(\sin(\omega\Delta t)/\omega)x_{s-1}^{b} \\
    x_{s}^{b} = x^{b}({t_{s-1}+\Delta t}) &= -\omega\sin(\omega\Delta t)x_{s-1}^{a}+\cos(\omega\Delta t)x_{s-1}^{b}.\\
    \end{aligned}
    \label{eq:SineSystem}
\end{equation} 
In our experiments, we let $\omega=0.04$ and $\Delta t=1$ to derive the true state $x_{t}^{a,\textrm{true}}$ at time step $t$. We assume that the measurement is the absolute value of the sum of $x_{t}^{\textrm{a,true}}$ and zero-mean Gaussian noise at the discrete times $t=1,\ldots,500$:
\begin{equation}
 y_{t}^{*} = \left|x_{t}^{a,\textrm{true}} + \gamma\right|, \quad \gamma \sim \mathcal{N}(0,\sigma_{obs}),
\end{equation}
where $\sigma_{obs}=0.1$ is the observation error standard deviation. The nonlinear observation operator complicates the inference problem since different values of the states can give the same measurement value. The noisy observations do not provide sufficient information to construct a unique estimate.
In this experiment, we assume that the true data-generating model is not known; instead, our forecast model consists of a one-dimensional random walk, paired with the correct nonlinear measurement model:
\begin{equation}
    \begin{aligned}
    & x_{t}^{a} = x_{t-1}^{a} + \epsilon,     \quad && \epsilon \sim \mathcal{N}(0,0.1), \\
    & y_{t} = \left|x_{t}^{a} + \gamma\right|,  && \gamma \sim \mathcal{N}(0,\sigma_{obs}),
    \end{aligned}
\end{equation}
where initial ensemble samples are drawn from a Gaussian prior $x_{1}^{a}\sim\mathcal{N}(0,0.2)$. In this imperfect setting, the effect of model error in the one-dimensional state equation makes it more difficult for the linear smoother to identify the correct bimodal structure of the posterior density.

\subsubsection{Experimental results}

\begin{figure}[!ht]
  \centering
  \includegraphics[width=\textwidth]{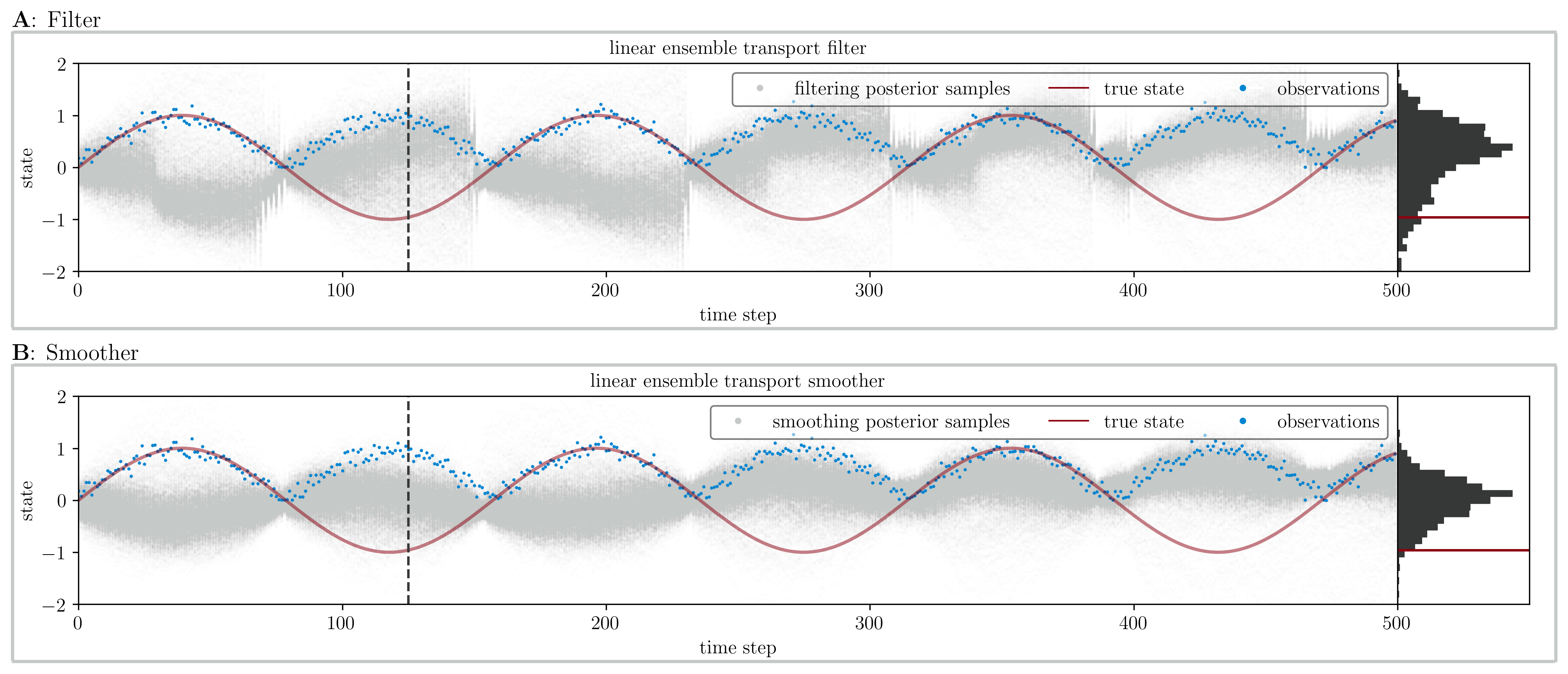}
  \caption{A combination of a filter (A) and a smoother (B) with linear updates fails to track multimodal distributions.}
  \label{fig:bimodal_sine_linear}
\end{figure}
\begin{figure}[!ht]
  \includegraphics[width=\textwidth]{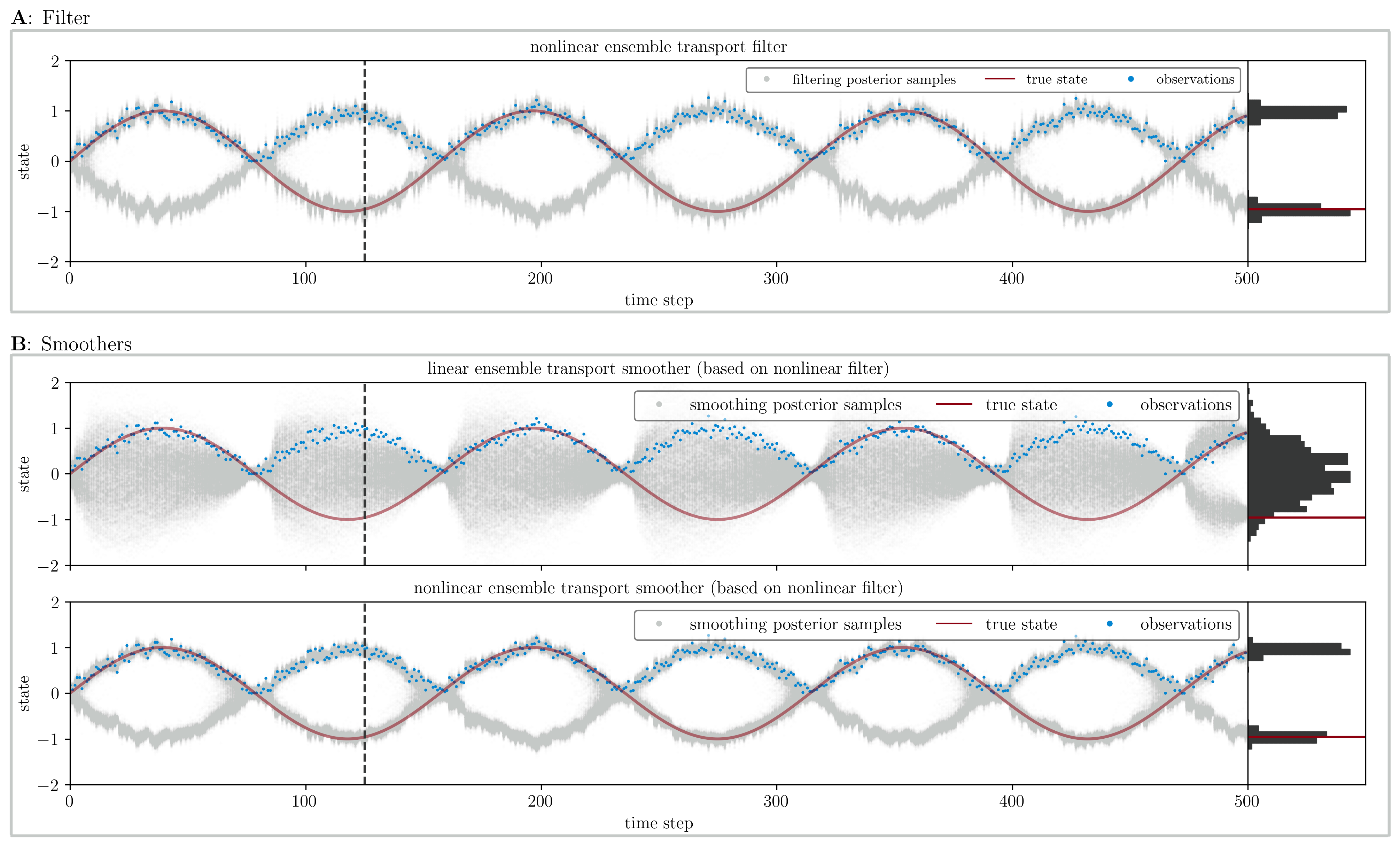}
  \caption{Filters with linear updates can track multimodal distributions (A). Using nonlinear maps for the smoother remains important, or the multi-modality recovered by the filter might be lost (B).}
  \label{fig:bimodal_sine_nonlinear}
\end{figure}
Figure~\ref{fig:bimodal_sine_linear} illustrates the performance of the linear single pass ensemble transport smoother for the random walk forecast model. The samples generated in the filtering and smoothing procedures are indicated with clouds of grey particles on the time series plots. Histograms showing the distribution of sampled values at the time indicated by the vertical dashed line are provided on the right end of each panel, with the true value indicated by a red line.  The linear filter shown in Figure~\ref{fig:bimodal_sine_linear}A attempts to track the mean and standard deviation of the correct posterior. The balance between both modes is unstable and the algorithm eventually snaps to one of the branches. The linear backward smoother (Figure~\ref{fig:bimodal_sine_linear}B) flattens out these patterns but similarly fails to recover the two separate modes.

We employ a nonlinear single pass filter that uses radial basis functions (RBFs) and cross terms. More detail on the exact parameterization is provided in Appendix~\ref{sec:AppendixC}.
This nonlinear filter is able to track both modes simultaneously (Figure~\ref{fig:bimodal_sine_nonlinear}A). 
An important feature of our ensemble transport smoothers is that we can freely adapt the complexity of the map components and thus improve the accuracy of the inference steps. In consequence, it is possible to combine a nonlinear EnTF with either a linear EnTS or a nonlinear EnTS.

Contrasting both options provides further insight into the smoothing performance (Figure~\ref{fig:bimodal_sine_nonlinear}B). As may be expected, a nonlinear EnTS preserves the bimodal structure of the nonlinear filter marginals. The linear EnTS, however, performs markedly worse: while it  preserves the filter's bimodal structure during the first part of the backwards pass (down to about $t \approx 475$), it cannot separate the two modes again once they merge at a zero crossing, affecting all subsequent smoothing steps. This demonstrates some of the risks in applying linear algorithms to non-Gaussian distributions.

If the structure of the true model in Equation~\ref{eq:SineSystem} is known but the initial states are  uncertain, we can use the discretized state equations to perform the probabilistic forecasts required by a recursive ensemble transport smoother. 
In this case, a filter that uses nonlinear updates is initially required to identify the bimodal structure from the non-negative  measurements, but we observe that a linear smoothing algorithm subsequently suffices to preserve the bimodality.
The detailed formulation and results for this ``identical twin model experiment'' is presented in Appendix~\ref{sec:AppendixD}.

\subsection{Lorenz-63}\label{subsec:L63}

Next, we examine the performance of the backward nonlinear EnTS applied to the Lorenz-63 model \citep{Lorenz1963DeterministicFlow}. This dynamical system has three scalar states $\x(t) = (x^a(t),x^b(t),x^c(t))$ that evolve in time according to the coupled ODEs
\begin{equation}
    \frac{d x^{a}}{d t} = \sigma(x^{b} - x^{a}), \quad \frac{d x^{b}}{d t} = x^{a}(\rho - x^{c})-x^{b}, \quad 
    \frac{d x^{c}}{d t} = x^{a}x^{b} - \beta x^{c},
    \label{eq:L63_scalar}
\end{equation}
where $\sigma,\beta,\rho \in \mathbb{R}$ are specified model parameters. For the parameter values $\sigma=10$, $\beta=\frac{8}{3}$, and $\rho=28$, which we select in our experiments, the model displays chaotic dynamics. In our implementation, we assimilate observations every $\Delta t = 0.1$ time units. We obtain forecast samples by simulating the dynamics over this interval using a fourth-order Runge-Kutta scheme, with integration step size $\Delta t = 0.05$. The prior for the initial state is set to the standard multivariate Gaussian distribution. The initial ensemble and initial synthetic true state $\X_{1}^{\textrm{true}}$ are drawn from this prior. Synthetic observations are subsequently sampled as $\y_{s}^{*} \sim \mathcal{N}(\X_{s}^{\textrm{true}},4\mathbf{I})$ for $s > 1$, following the setup of \citet{Lei2011ABickel}. We evaluated the single-pass EnTS over $1000$ time-steps following a $250$ timestep spin-up period with an empirical EnKF. The forecast and observation models can be concisely written in vector form as:
\begin{equation}
    \begin{aligned}
    &\x_{s} = \mathbf{\operatorname{L63}}(\x_{s-1};\sigma,\beta,\rho)\\
    &\y_{s} = \x_{s} + \boldsymbol{\epsilon}, \quad \boldsymbol{\epsilon} \sim \mathcal{N}(\mathbf{0}, 4\mathbf{I}),
    \end{aligned}
    \label{eq:L63_model}
\end{equation}
where $\operatorname{L63}$ is a vector operator that represents the discrete-time state update of Equation~$\ref{eq:L63_scalar}$ over $\Delta t = 0.1$. We test the smoothing algorithms for different ensemble sizes $N$ between $50$ and $1000$, and different transport map complexities. As is common in ensemble filtering, especially for small ensemble sizes, we also consider three different inflation factors, $\gamma \in \{0,0.1,0.2\}$. We realize inflation as
\begin{equation}
    \begin{aligned}
    &\widehat{\x}_{s} = \overline{\x}_{s} + \sqrt{1+\gamma}(\x_{s} - \overline{\x}_s) \\
    &\widehat{\y}_{s} = \widehat{\x}_{s} + \boldsymbol{\epsilon}, \quad \boldsymbol{\epsilon} \sim \mathcal{N}(\mathbf{0}, 4\mathbf{I}),
    \end{aligned}
    \label{eq:L63_model_inflated}
\end{equation}
where $\overline{\x}_{s}$ is the ensemble mean of $\x_{s}$, and $\widehat{\x}_{s}$ and $\widehat{\y}_{s}$ are the inflated state and prediction ensemble members for time step $s$. We use the inflated samples only to estimate the transport maps used for filtering, and otherwise use $\x_{s}$ and $\y_{s}$ as the inputs for the composite map. We avoid inflation for smoothing pass as suggested in~\citet{Raanes2016OnSmoother}. In addition to inflation, we add an $L^2$ regularization penalty on the map coefficients to the objective function in Equation~\ref{eq:objective_function} for learning the map. The regularized objective function is given by
\begin{equation}
\widetilde{\mathcal{J}}_{k}(S_{k}) = \widehat{\mathcal{J}}_{k}(S_{k}) + \lambda\sum_{j=1}^{P_{k}} c_{j}^2,
\label{eq:objective_function_l2reg}
\end{equation}
where $P_{k}$ is the number of coefficients in the $k$th map component $S_{k}$ (see Equation~\ref{eq:separable_map}).
We consider regularization parameters $\lambda \in \{0,0.5,1,1.5,2\}$. The map parameterization used in this example is based on separable maps (see Section~\ref{subsec:monotonicity}) using Hermite functions for the nonmonotone terms and integrated RBFs for the monotone terms. More detail is provided in Appendix~\ref{sec:AppendixE}. In our experiments, we examine the smoothing performance for each ensemble size when varying the order for the basis functions in the map components from one (i.e., affine maps) to five. As described in \citet[Appendix E]{Ramgraber2022underUpdates}, we capitalize on the independent observation errors by assimilating the observation components $y_{t}^{a}$, $y_{t}^{b}$, and $y_{t}^{c}$ one at a time during the filtering pass, which yields lower-dimensional and sparser maps.

To gain further insight into the performance of the nonlinear backward EnTS, we compare our results to an iterative Ensemble Kalman Smoother (iEnKS) \citep{Bocquet2014AnSmoother} using the Dapper toolbox \citep[ver.~1.3.0: ][]{Raanes2016OnSmoother}. We implemented the iEnKS with a multiple data assimilation (MDA) formulation and compared two update rules: perturbed observations (PertObs), which are closer to our EnTS implementation, and a square-root (Sqrt) formulation, based on semi-empirical identities assuming Gaussian errors with known covariance matrix. We explored all combinations of different ensemble sizes $N \in [5,10,25,50,100]$, inflation factors $\gamma \in [1.02,1.04,1.07,1.1,1.2]$, and smoothing lags $L \in [1,2,3,4,5,19,15,20,25,30]$. We then repeat all simulations for $10$ different random seeds, and set the maximum number of iEnKS iterations to $\operatorname{nIter}=10$. 

\subsubsection{Experimental results}

We simulate combinations of (i) ensemble size, (ii) map order/polynomial degree, (iii) inflation factor, and (iv) regularization factor, and repeat all simulations ten times with different random seeds. As a performance metric, we report the time-averaged root mean square error (RMSE) in our ensemble mean estimate of the state. We average both metrics across all random seeds, then report the optimal inflation/regularization factor combination over the full assimilation window for every ensemble size and map order combination. 
\begin{figure}
  \centering
  \includegraphics[width=\textwidth]{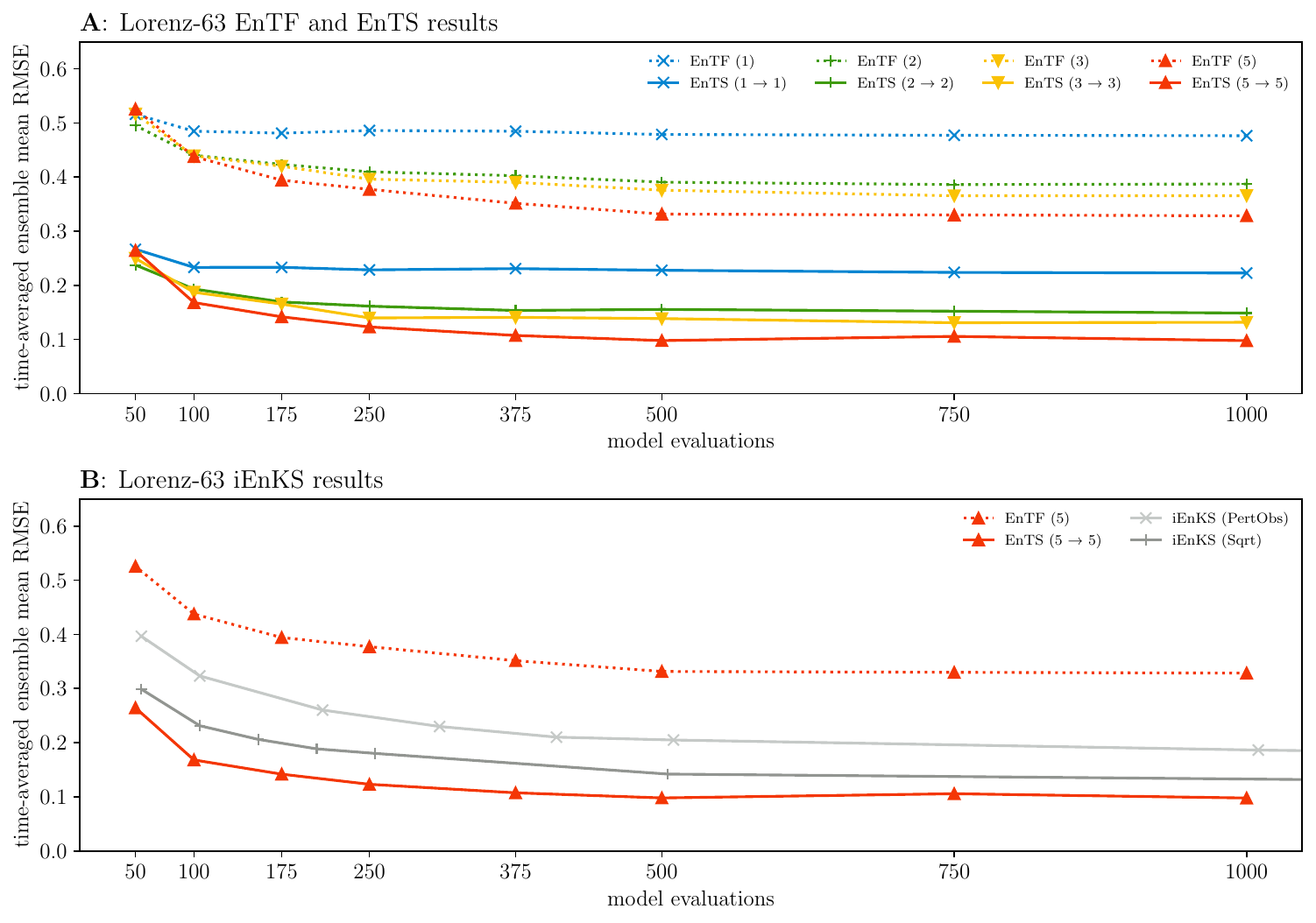}
  \caption{RMSE for ensemble transport filters and smoothers of different complexity and ensemble size, calculated relative to the synthetic true state. For the EnTF and EnTS, the number of model evaluations corresponds to their ensemble size. Smoothers outperform filters, and both EnTF and EnTS improve with greater nonlinear complexity provided the ensemble size is large enough (A). An EnTS of sufficient complexity can outperform the iEnKS over a fixed number of model evaluations (B). The label ``smoother~($m$~$\rightarrow$~$n$)'' refers to an order-$n$ EnTS combined with an order-$m$ EnTF, while ``EnTS~($m$)'' refers to an order-$m$ EnTF.}
  \label{fig:L63_results}
\end{figure}

Results are illustrated in Figure~\ref{fig:L63_results}. A first observation is that the smoothers yield substantial improvements over the filters across all ensemble sizes or map orders considered. As ensemble size increases, the RMSE reduces for both filters and smoothers. The amount of improvement, however, varies for different map orders: the RMSE of linear EnTF and EnTS saturate quickly, flattening out past an ensemble size of $N=100$. This is due to the structural limitations of simpler transport maps, with insufficient expressiveness to capture the target distribution. Their nonlinear counterparts provide further improvements over the linear algorithms, but greater complexity demands larger ensemble sizes to realize its potential. At the same time, in the small ensemble size regime linear smoothers outperform their nonlinear counterparts. This suggests that if ensemble sizes are not sufficiently large, simpler transport maps can be a safer option, an instance of the underlying bias-variance tradeoff. 

Figure~\ref{fig:L63_results}B shows the results of the iEnKS.
Each iEnKS simulation demands $N\cdot(\operatorname{nIter}\cdot L+1)$ model evaluations. As the EnTS is not an iterative algorithm, we report the iEnKS's computational demand in terms of the total number of model evaluations rather than ensemble size. Our results are encouraging, and suggest that the nonlinear EnTS can match and even surpass the iEnKS's performance in this system. Eventually---after a significantly larger number of model evaluations than shown in the figure---the iEnKS reaches RMSEs similar to the $N=1000$ order-5 EnTS; this occurs at an equivalent cost of $3000$ (iEnKS-Sqrt) and $30000$ (iEnKS-PertObs) model evaluations. 

The continuous rank probability score (CRPS)  between the ensemble and the true state \citep{Gneiting2007ProbabilisticSharpness,Brocker2012EvaluatingScore} provides useful information on the accuracy of the posterior ensemble spread. The CRPS values for our L63 experiment, not shown here, reveal the same patterns as the RMSE results plotted in Figure~\ref{fig:L63_results}, indicating that the ensemble spreads for smoothers are consistently better than for filters and  that nonlinear smoother spreads improve as their order is increased. Also, the transport smoothers give CRPS scores that are somewhat better than the iEnKS scores for comparable computational effort.

The greater efficiency of the square-root formulation \citep[e.g., ][]{Evensen2019EfficientMatching} relative to the perturbed observation formulation \citep[e.g., ][]{Emerick2013EnsembleAssimilation} is based on the former's use of analytical error identities. These identities partially lift the burden of estimating the error statistics from the ensemble, hence improving performance, but require that the observation errors are specifically additive and Gaussian. The iEnKS-PertObs and EnTS, by contrast, rely on realizations of the observation errors, and can thus accommodate a wider range of stochastic observation models.\footnote{Beyond the fully sample-driven EnTS algorithms considered in this study, there also exist nonlinear transport analogues of square-root filters, called deterministic map filters \citep{Spantini2022CouplingFiltering}. In principle, these could similarly be extended to smoothing.} 
Additional detail about the iEnKS results is provided in Appendix~\ref{sec:AppendixF}. 

\subsubsection{Computational costs}\label{subsubsec:computational_demand}

\begin{figure}
  \centering
  \includegraphics[width=\textwidth]{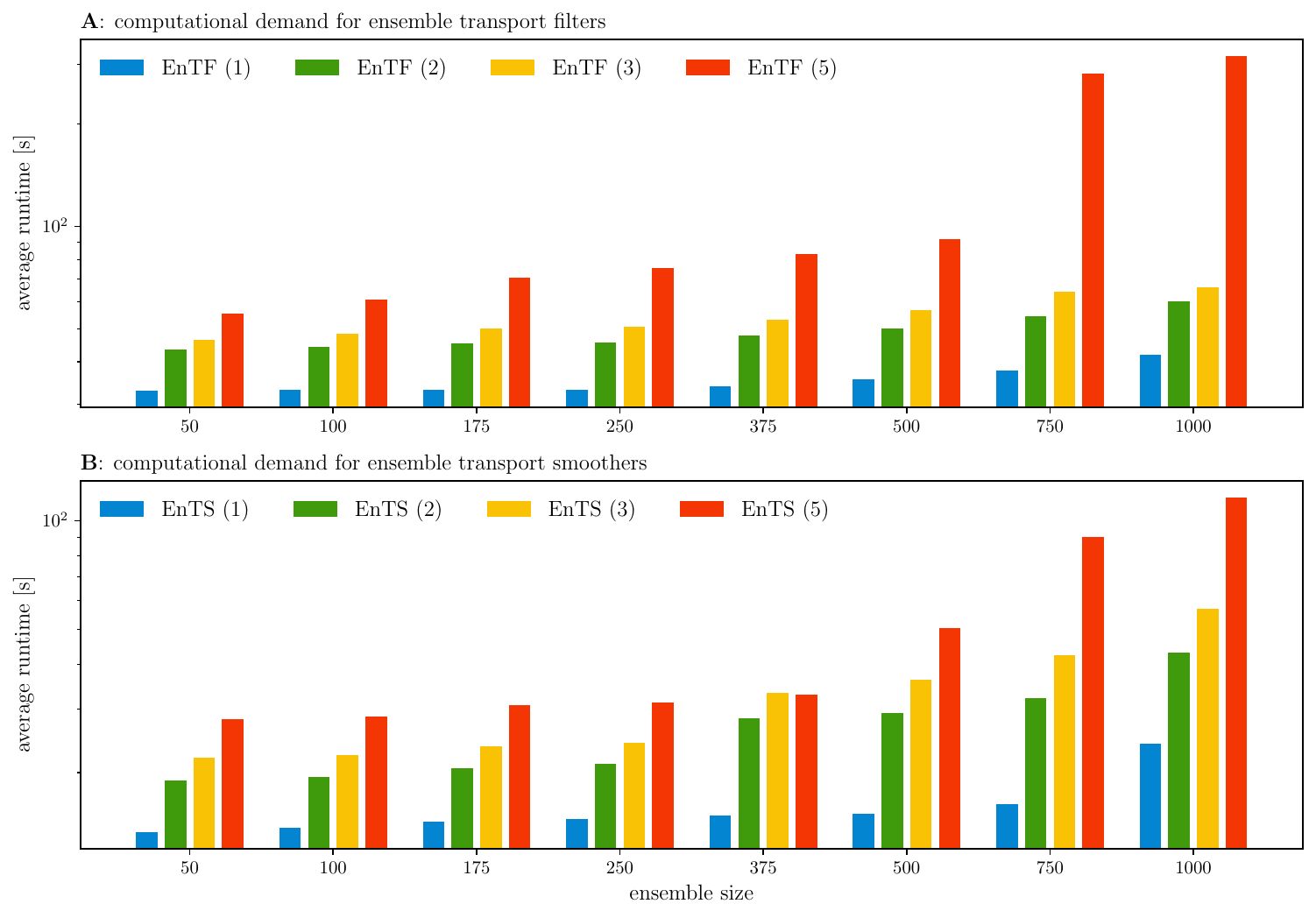}
  \caption{Wall clock time for a full EnTF filtering (A) and backward EnTS smoothing (B) pass for the Lorenz-63 example. The runtime is reported for a time series length of $t=1000$, and for different ensemble sizes $N$ and orders of map complexity (from $1$ to $5$, each corresponding to a different color).}
  \label{fig:L63_time_demand}
\end{figure}

Figure~\ref{fig:L63_time_demand} illustrates the computational costs of the EnTF and EnTS algorithms for varying levels of map complexity. 
The reported runtimes are for a full filtering or smoothing pass over a time series with $t=1000$ steps, averaged over multiple runs. We observe an increase in runtimes for both algorithms with larger ensemble sizes $N$ and higher order parameters, which define the map complexity. We observe relatively constant runtimes over small to moderate ensemble sizes (i.e., $N \leq 375$), likely due to computational overhead of our implementation. We note that overall runtimes remain relatively low, as compared to the typical computational costs of model evaluations in applied data assimilation problems, 
which are unaffected by the choice of inference routine.

\subsection{Lorenz-96}\label{subsec:L96}

We now investigate the performance of a nonlinear backwards EnTS for the Lorenz-96 system~\citep{Lorenz1995Predictability:Solved,Lorenz1998OptimalModel}. This system was originally intended to simulate the evolution of a generic scalar quantity, such as temperature, that is transported in a fluid over time and through a circular (e.g., constant latitude) spatial region. The system is defined in a spatially discretized form by the set of coupled ordinary differential equations
\begin{equation}
    \frac{d x_{j}}{d t} = (x_{j+1} - x_{j-2})x_{j-1} - x_{j} + F, \quad j=1,\dots,J,
    \label{eq:Lorenz-96}
\end{equation}
where $j$ represents one of $J$ equally-spaced grid points on the circle. Note that each state depends asymmetrically on its neighbours and has periodic boundary conditions, 
i.e., $x_{J+1}=x_{1}$, $x_{0}=x_{J}$, and $x_{-1}=x_{J-1}$. Indices increase in the positive direction and decrease in the negative direction. The nonlinear terms account for advection, the linear term accounts for dissipation, and the variable $F$ is a forcing term. In this study, we set $F=8$, which results in chaotic dynamics, and choose $J=40$. We set the inter-observation time to $\Delta t=0.4$ and integrate the dynamics using a fourth-order Runge-Kutta scheme with a time step of $0.01$. We assume that the system evolves without model error, and we observe every other state with independent Gaussian additive errors with a standard deviation of $\sigma_{obs}=0.25$. We test the performance of the backwards EnTS for different ensemble sizes $N\in[50,100,135,175,250,375,500]$, inflation factors $\gamma\in[1.0,1.025,1.05,1.1,1.2]$, regularization factors $\lambda\in[0,1,3,7,10,15]$, and map orders $O\in[1,2,3,5]$. As for Lorenz-63, inflation was only used for the EnTF. Detail on the map parameterization is provided in Appendix~\ref{sec:AppendixE}. 

\subsubsection{Localization}

For high-dimensional systems such as Lorenz-96, it is common to localize the states $\x$ and observation predictions $\y$ in space, reflecting the fact that two states far removed from each other in terms of indices can often be considered independent. In this study, we propose a new, general method to derive empirical localization lengths from a preceding linear EnTF simulation. This is achieved in several steps:

\begin{enumerate}
    \item First, we run an EnKF with a large ensemble size ($N=1000$) to obtain forecast state and observation samples $(\Y_{s},\X_{s}) \sim p(\y_{s},\x_{s}|\y_{1:s-1}^{*})$ for the filter and state samples $(\X_{s+1},\X_{s}) \sim p(\x_{s+1},\x_{s}|\y_{1:s}^{*})$ for the smoother, at varying times $s$. We standardize each of these ensembles to have unit marginal variances.
    \item Next, we use the ensembles to construct sample estimates of the covariance matrices of these joint distributions at each time, and take their element-wise absolute values.
    \begin{enumerate}
        \item For smoothing, we then average these matrices over time to obtain a mean covariance $\bar{\boldsymbol{\Sigma}}_{\X,\X}^{\textrm{smooth.}}\in\mathbb{R}^{80 \times 80}$.
        \item Similar to Lorenz-63 (Section~\ref{subsec:L63}), our Lorenz-96 filter assimilates every observation independently. For the assimilation of $y_{s}^{m,*}$, we arrange the states as $[y_{s}^{m},x_{s}^{m},x_{s}^{m-1},x_{s}^{m+1},x_{s}^{m-2},x_{s}^{m+2},\dots]$, with indices $m$ wrapping around. We write the state/observation covariance matrices using this ordering, and then average these matrices over each filtering update and each time step to obtain $\bar{\boldsymbol{\Sigma}}_{\Y,\X}^{\textrm{filter}}\in\mathbb{R}^{41 \times 41}$. 
    \end{enumerate} 
    \item We then invert these averaged covariance matrices to obtain averaged precision-type matrices $\bar{\mathbf{P}}_{\Y,\X}^{\textrm{filter}}$ and $\bar{\mathbf{P}}_{\X,\X}^{\textrm{smooth.}}$ for the filtering and smoothing updates, respectively.
    \item Finally, for each row of the precision-type matrices, we incrementally explore the entries on both sides of the diagonal and mark entries that are above an absolute precision threshold of $0.01$. If the entries on both sides fall below this precision threshold, we stop and proceed to the next row. We then average the (aligned) cell markings across all rows to derive asymmetric localization lengths in both state index directions. For the smoother, we also repeat this procedure for the $D$-th subdiagonal.
\end{enumerate}

\begin{figure}
  \centering
  \includegraphics[width=0.9\textwidth]{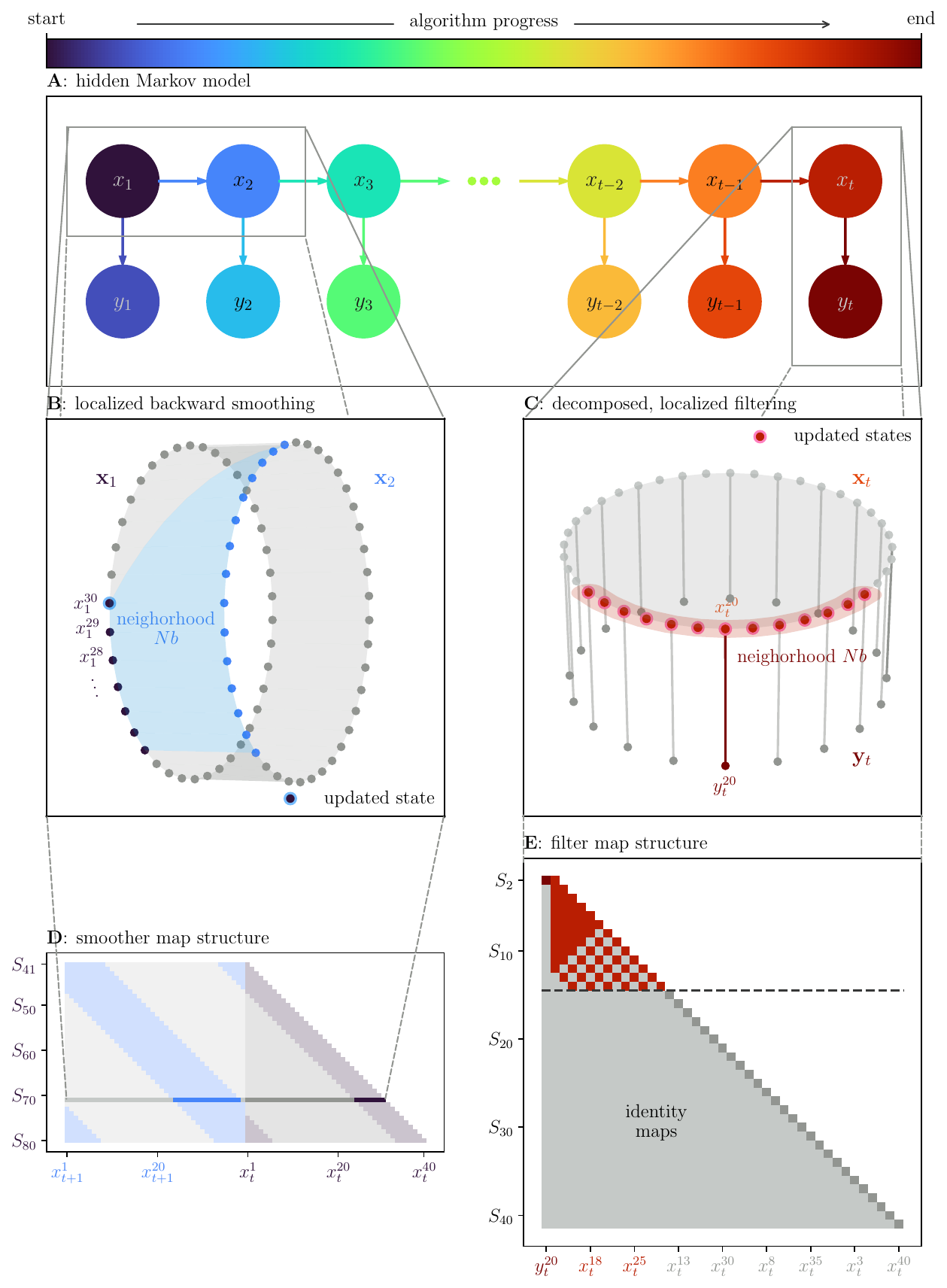}
  \caption{Graphical structures used for the Lorenz-96 simulation. Across-time dependencies of the state and observations are described by the standard hidden Markov model (A). Subdividing and localizing the smoothing (B) and filtering (C) operations can exploit sparse dependence structure \textit{within} the state, among the components of adjacent pairs of states, and between state components and their observations. Subplot B illustrates the dependencies (D) of one map component $S_{40+30}$ in the full map, subplot C shows the dependencies (E) for the decomposed filter's assimilation of observation $y_{t}^{20}$ in the full map. Grey blocks are sparsified away.}
  \label{fig:Lorenz-96}
\end{figure}

This procedure yields a symmetric state localization length of $6$ for the the individual filtering updates. For the backward smoother, localization lengths vary: Along the diagonal ($\X_{s}$), we find a symmetric state localization length of $6$. For the cross-dependence on the future states $\X_{s+1}$ along the subdiagonal, we find asymmetric localization lengths: $6$ indices in the negative direction, and $8$ indices in the positive direction. This asymmetry arises from the state equations (Equation~\ref{eq:Lorenz-96}), which depend on more states in the negative direction than in the positive direction. Illustrations of these localization patterns for filtering and backwards smoothing are provided in Figure~\ref{fig:Lorenz-96}. Note this approach differs from many other localization schemes in that it is \textit{precision-based} \citep{Nino-Ruiz2018AnEstimation} rather than covariance-based, thus relying on the decay of \textit{conditional dependence} rather a decay of \textit{dependence} or correlation. In a sense, our heuristic aims to discover the structure of the underlying undirected graphical model (under linear--Gaussian assumptions), and uses this information to remove variable dependencies.

\subsubsection{Experimental results}

\begin{figure}
  \centering
  \includegraphics[width=\textwidth]{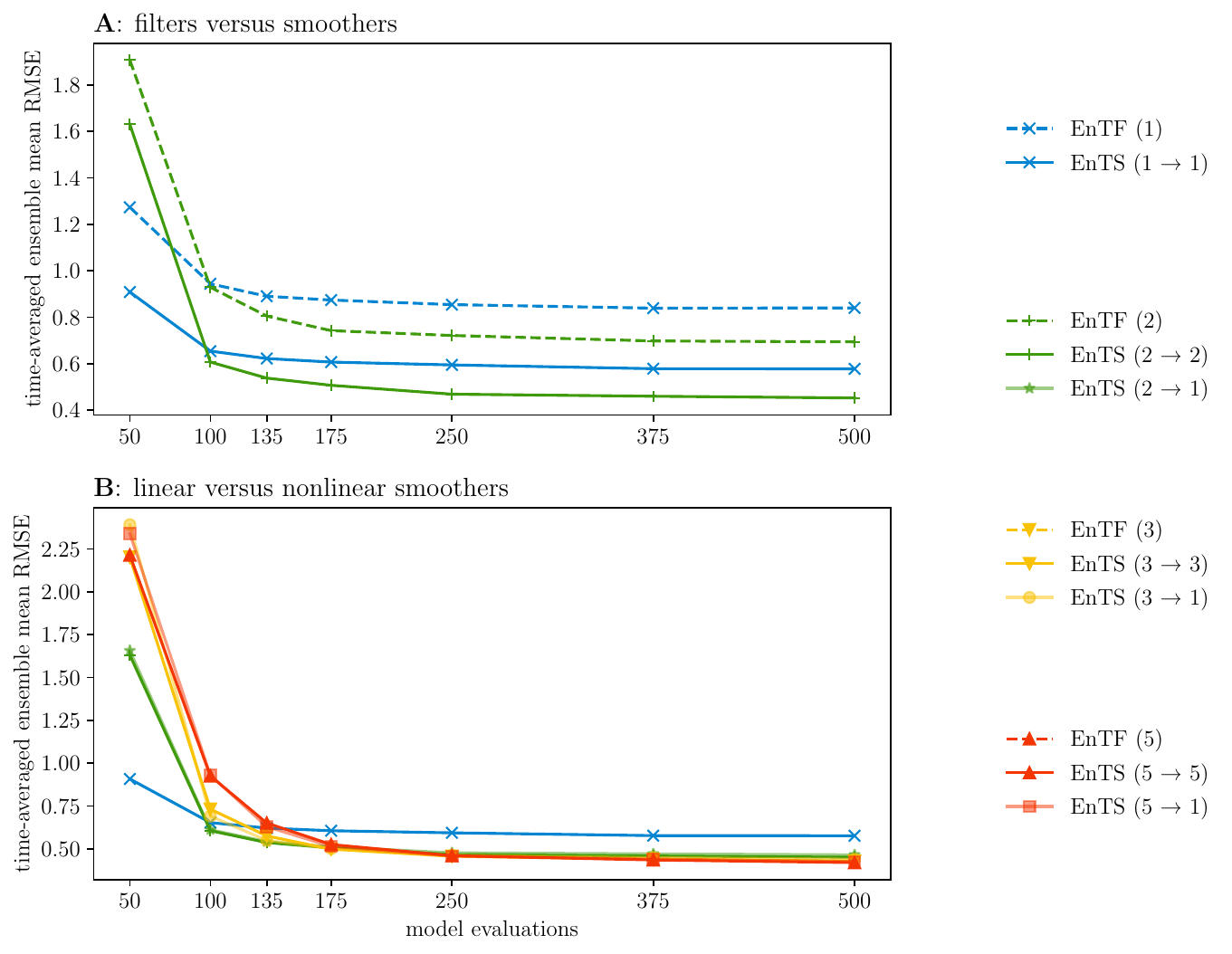}
  \caption{RMSE of ensemble mean estimates in the Lorenz-96 example. (A) Linear and quadratic EnTF and EnTS over a range of  ensemble sizes. Smoothers outperform filters, provided the ensemble size is sufficiently large. However, nonlinear transport maps do not seem to be required for the backward smoothing pass (B). Linear smoothing maps generally suffice. The label ``EnTS~($m$~$\rightarrow$~$n$)'' refers to an order-$n$ EnTS combined with an order-$m$ EnTF, while ``EnTS~($m$)'' refers to an order-$m$ EnTF.}
  \label{fig:L96_results}
\end{figure}

Results in terms of RMSE are illustrated in Figure~\ref{fig:L96_results}. Overall, higher degrees of nonlinearity in the maps result in better filtering and smoothing performance, with a sharp drop in error from order $1$ to order $2$, and diminishing returns for higher orders. This improvement seems to be mainly inherited from the filter, with nonlinear smoothers only providing negligible benefit over linear maps. Similar to the results for Lorenz-63, higher map orders must be supported by proportionally larger ensemble sizes. Analyzing the optimal results for each level of smoothing map complexity, we observed the best results for high $L^2$ regularization factors. As regularization serves as protection against overfitting, this suggests that much of the complexity in the nonlinear smoothing maps might be superfluous, resulting in unfavourable bias-variance trade-offs.

To develop further insight, we analyze the joint distributions $p(\x_{s},\x_{s-1}|\y_{1:s-1}^{*})$ underlying the backward smoothing operations. To this end, we ran a semi-empirical EnKF (see \citet{Ramgraber2022underUpdates}) with an ensemble size of $N=10000$. From this run, we extracted a joint analysis/forecast ensemble (i.e., $(\X_{s-1}^\ast, \X_{s})$) from each of the final $100$ time steps $s$. We then whitened each such ensemble by: (1) subtracting the ensemble mean, thus centering it on the origin; and (2) multiplying by the Cholesky factor of the inverse of the ensemble covariance, thus transforming the ensemble to have identity covariance. Together, these operations mimic the effect of an optimal linear transport map. If the ensemble members were distributed according to any multivariate Gaussian, their linearly whitened versions would be \textit{standard} Gaussian. Insofar as the ensembles are not jointly Gaussian, nonlinear transport maps are beneficial for smoothing. 

\begin{figure}[!ht]
  \centering
  \includegraphics[width=\textwidth]{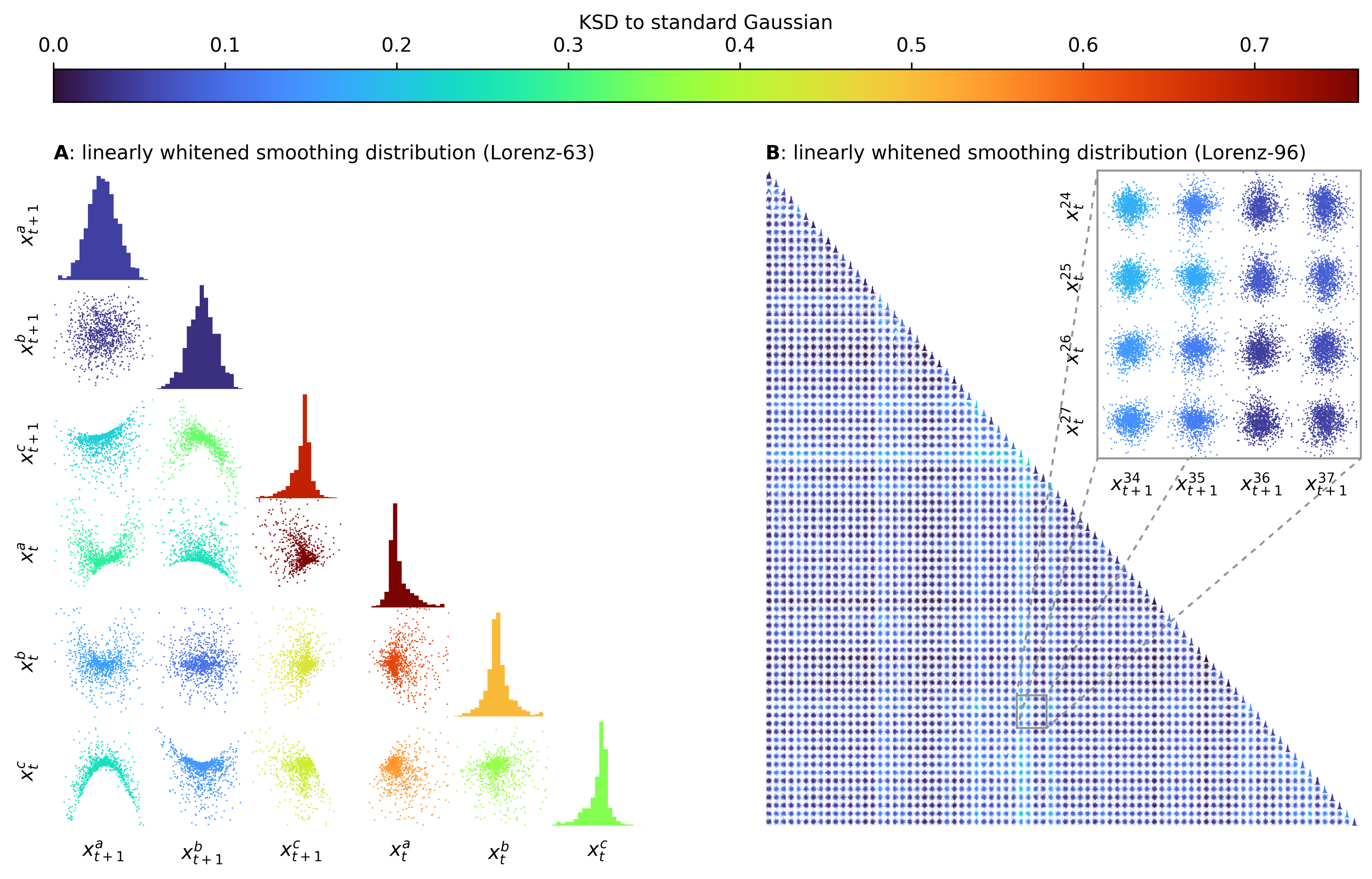}
  \caption{Univariate and bivariate samples from the whitened joint distributions $p(\x_{s},\x_{s-1}|\y_{1:s-1}^{*})$ that are used to construct smoothing maps for (A) Lorenz-63 and (B) Lorenz-96. The samples are linearly transformed to have zero mean and identity covariance before plotting. The color denotes the kernelized Stein discrepancy (KSD) between the univariate (histogram) or bivariate (scatter plot) marginal distributions and a standard Gaussian reference.
  Lorenz-63 states retain substantial non-Gaussianity after whitening, whereas the Lorenz-96 states do not. The ensemble was sub-sampled $10:1$ for the KSD estimation.}
  \label{fig:Lorenz-96-forensics}
\end{figure}

We quantified the degree of non-Gaussianity in each of the bivariate marginals of the linearly whitened ensembles by computing the \textit{kernelized Stein discrepancy} (KSD)~\citep{Liu2016ATests} to a standard Gaussian distribution of dimension two, using an inverse multiquadric kernel with median distance bandwidth.
Figure~\ref{fig:Lorenz-96-forensics} illustrates the joint distributions $p(\x_{s},\x_{s-1}|\y_{1:s-1}^{*})$ with the largest average KSD among the last $100$ filtering timsteps $s$, for both Lorenz-63 (Figure~\ref{fig:Lorenz-96-forensics}A) and Lorenz-96 (Figure~\ref{fig:Lorenz-96-forensics}B). To aid the visual interpretation of these results, we have colored each subplot according to its respective KSD: deep blue corresponds to low KSDs, closely matching a standard Gaussian; deep red identifies the strongest departures from Gaussianity. We observe that the joint distributions involved in Lorenz-63 retain substantial non-Gaussianity after linear whitening, and consequently benefit from nonlinear transport maps during the smoothing pass. The joint filtering--forecast distribution for Lorenz-96, on the other hand, is already very close to standard Gaussian, and thus does not stand to benefit from nonlinear transport maps to the same degree. 

Recall that nonlinear transport maps allow us to approximate non-Gaussian target distributions, but if the target already is sufficiently Gaussian, nonlinear terms may only increase the variance of the map estimator without providing a reduction in bias. This suggests that the degree of Gaussianity and thus the required complexity of the transport maps may vary not only with the system dynamics themselves, but also between the filtering and smoothing passes. This observation is in line with the findings of \citet{Morzfeld2019GaussianAssimilation}.
\begin{figure}[!ht]
  \centering
  \includegraphics[width=\textwidth]{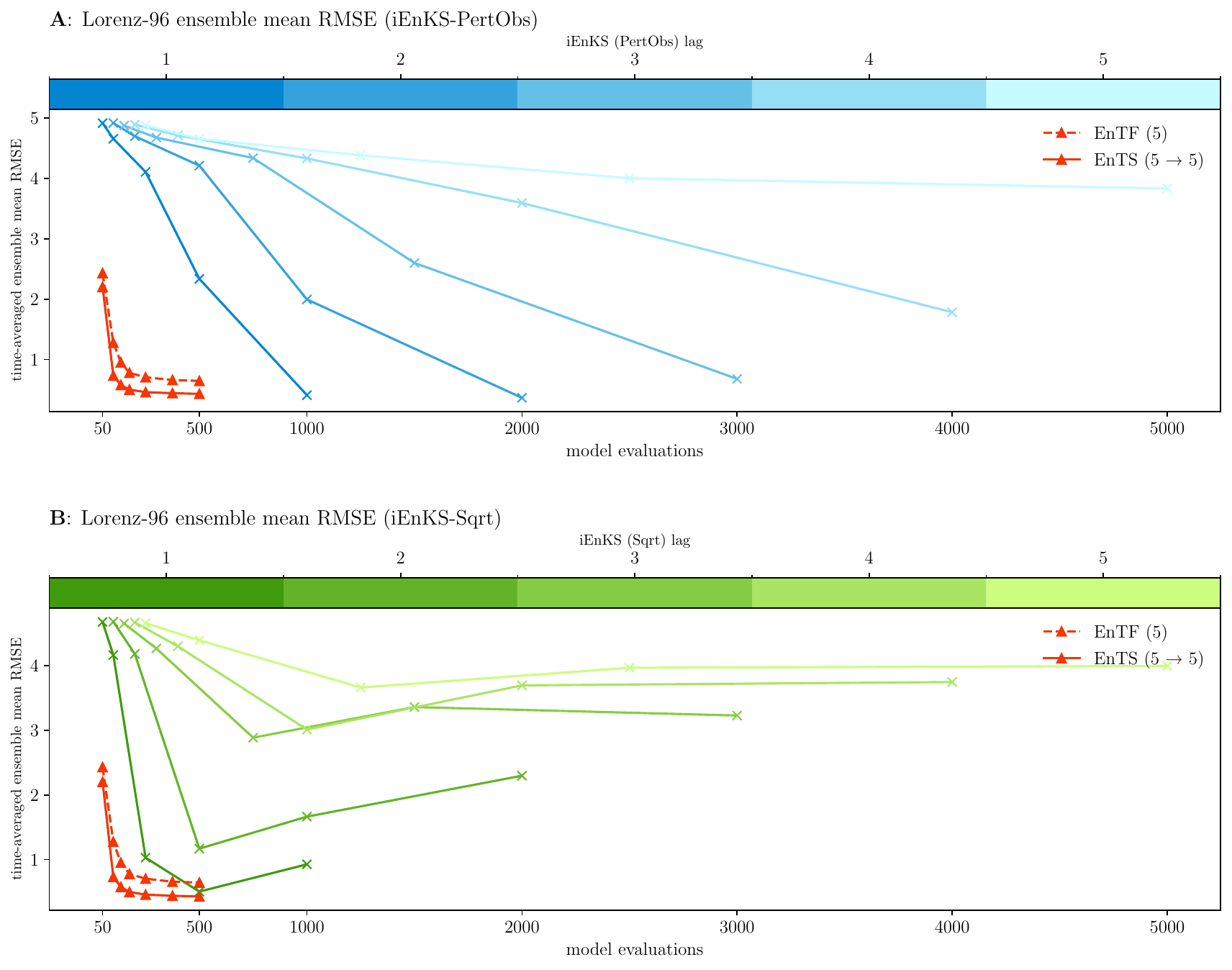}
  \caption{Lorenz-96 example. RMSE obtained with the iEnKS-PertObs (A) and iEnKS-Sqrt (B) over different lags and model evaluations, compared to results of the order-$5$ EnTS.}
  \label{fig:L96_results_iEnKS_vs_EnTS}
\end{figure}

Results of the iEnKS reference simulations are provided in Figure~\ref{fig:L96_results_iEnKS_vs_EnTS}. We preface the discussion of these results by noting that the iEnKS implementation we used does not at present support localization \citep[][v.~1.3.0]{Raanes2018DataDapper}, although the framework can support it in principle \citep{Bocquet2016LocalizationSmoother}. With this in mind, the iEnKS results reflect our findings in Lorenz-63: the iEnKS can match the smoothing performance of the EnTS, but at a higher equivalent cost of model evaluations. 

A curious observation is the degradation of iEnKS's performance with higher lags. This phenomenon is also noted in the literature \citep{Fillion2018Quasi-staticSmoother}, where it is attributed to the emergence of local minima in the objective function. While we did not investigate this phenomenon in greater detail, we would add that the iEnKS implements a dense smoothing update, and thus is likely to suffer from spurious updates to a greater degree than an equivalent backward smoothing formulation (see \citet{Ramgraber2022underUpdates}).

\section{Discussion}\label{sec:discussion}
\subsection{Conclusions}

\paragraph{\textbf{Bayesian inference}} Ensemble transport methods provide a flexible and efficient way to solve Bayesian inference problems, by identifying a map that transforms samples from the joint distribution of states and observations to samples from a tractable, user-specified, reference distribution. Components of this map can then transform reference samples into realizations of the joint distribution or any of its conditionals; the latter include the posterior distributions arising in Bayesian filtering and smoothing. Composing these operations together yields a transformation from a given joint distribution to a desired conditional.

\paragraph{\textbf{Ensemble transport smoothing}} The nonlinear ensemble transport smoothers (EnTS) developed in this paper apply transport ideas to smoothing problems of interest in geophysical applications. The EnTS combines the convenience and computational efficiency of classical recursive estimators, such as the ensemble Kalman smoother, with the generality of importance sampling methods---in particular, the ability to consistently handle non-Gaussianity. At each step of the recursive smoothing procedure, the EnTS uses a nonlinear triangular transport map (an approximation of the true Knothe--Rosenblatt rearrangement) to condition the relevant state ensemble on new measurements, or on updated realizations of a neighboring state.
 
\paragraph{\textbf{Sparse triangular transport maps}} The structure of a triangular transport map used for Bayesian inference reflects a chosen ordering of variables, which in turn corresponds to a specific factorization of the joint prior density of states and observations. In the presence of conditional independence, we can simplify this factorization, and sparsify the triangular map accordingly.
Applying this idea to state-space models, particular variants of the EnTS emerge from different orderings of the states \citep{Ramgraber2022underUpdates}. Here we focus on ensemble transport smoothers that use backward orderings appropriate for growing window multi-pass, fixed lag multi-pass, and fixed window single-pass applications.

\paragraph{\textbf{Adaptive complexity}} The expressiveness of a triangular map, i.e., the complexity of its parameterization, can be adapted to the demands of the target distribution. While linear maps may suffice for nearly Gaussian target distributions, more complex distributions demand nonlinear maps. This is relevant in smoothing, which includes filtering and smoothing operations that may not require the same level of complexity. For instance, in Lorenz-96 (see Section~\ref{subsec:L96}), we found it advantageous to use a nonlinear map in the filtering step and to keep the smoothing step linear. In applications where the relevant distributions at both stages are distinctly non-Gaussian, better performance is obtained by using nonlinear maps for both filtering and smoothing. The ability to tailor the map configuration to the application is one of the unique advantages of an ensemble transport approach.

\paragraph{\textbf{Localization}} Transport methods provide a convenient framework for incorporating time and space localization in data assimilation problems. In particular, localization takes the form of imposing \textit{sparse} variable dependence on the transport map estimated from an ensemble, and hence an ansatz of conditional independence. Such localization structure can be learned adaptively. In our Lorenz-96 experiment (Section~\ref{subsec:L96}), our transport smoothing algorithm implements localization by identifying conditional independence from time and space-averaged precision matrices. This method is one example of a general transport-oriented approach to localization that can be applied in many different situations.

\paragraph{\textbf{Experiment: bimodal sine}} Nonlinear filtering and smoothing updates may be required to obtain accurate probabilistic assessments, even in deceptively simple problems. Our first smoothing experiment combines a misspecified autoregressive model, a nonlinear measurement operator, and Gaussian initial states. The resulting system exhibits periodic bimodality in its filtering and smoothing distributions. Nonlinear maps are required in both the filtering and smoothing operations to recover the correct posterior. This experiment suggests that the need for nonlinear smoothing updates depends on both the accuracy of the forecast model and the degree of non-Gaussian behavior encountered.

\paragraph{\textbf{Experiment: Lorenz-63}} Our experiments with the chaotic three-state Lorenz-63 model demonstrate the benefits of using higher-order polynomial basis functions in the transport map when the system behavior is highly non-Gaussian. The non-Gaussian nature of the forecast distributions requires nonlinear updates for both filtering and smoothing. The performance of the EnTS compares favorably to that of an iterative ensemble Kalman smoother (iEnKS), giving comparable or better accuracy and ensemble spread with less computational effort.

\paragraph{\textbf{Experiment: Lorenz-96}} Our experiments with a chaotic 40-dimensional Lorenz-96 model demonstrate the benefits of adapting map complexity to different stages of the inference problem.
Using localization and a map comprised of polynomial basis functions, we found that the filtering pass benefits from nonlinear maps, whereas linear maps suffice for the smoothing pass, due to near-Gaussianity of the relevant filtered/forecast distributions. As in Lorenz-63, the performance of the EnTS compares favorably to that of an iEnKS, giving comparable estimation performance with less computational effort.

\subsection{Limitations and technical details}

Due to differences in their implementation, we could not reproduce identical setups of the EnTS for the iEnKS, which might affect the comparisons explored in Section~\ref{sec:experiments}. For instance, the nonlinear observation model of the bimodal sine example could not be easily realized with the iEnKS; hence its absence in this scenario. Furthermore, localization for the iEnKS is not supported in the Dapper toolbox at time of writing, although it is possible in principle \citep{Bocquet2016LocalizationSmoother}. Consequently, we did not use localization for our iEnKS reference simulations in the Lorenz-96 experiments.

As far as computational cost is concerned, the EnTS had the greatest computational demand for the realization of a smoothing update among the algorithms we considered. We note, however, that this cost only pertains to the smoothing update itself. In most systems, the limiting factor is the number of model evaluations, which remains unaffected by the chosen inference routine (with the exception of the iEnKS, whose iterative updates depend on additional model evaluations). For Lorenz-63, the average computational demand on an Intel® Core™ i7-7700K CPU varied between $12$ ms (order $1$) and $20$ ms (order $5$) for a single EnTF filtering operation, and between $14$ ms (order $1$) and $24$ ms (order $5$) for the corresponding EnTS update operation.

\subsection{Outlook}

Triangular transport methods hold substantial potential for Bayesian inference, and constitute a versatile tool for filtering and smoothing alike. While we explored several useful features of nonlinear transport maps, many further research avenues remain: 
\begin{itemize}
    \item \textbf{Dimension reduction} techniques \citep{Jollife2016PrincipalDevelopments,Scheidt2018QuantifyingSystems,Solonen2016OnFilters,Baptista2022Gradient-basedPerspective}, for both states and observations, lower the number of input variables to the map $\SKR$ and the number of map components $S_{k}$ that need to be estimated. This reduces the ensemble size required to accurately estimate the state and the computational demand of nonlinear filtering and smoothing algorithms; see an application with vortex flow models  in~\cite{le2022low}.
    Ensemble Kalman-type algorithms commonly reformulate their updates in low-dimensional ensemble subspaces, thereby performing implicit dimension reduction. Restricting the updates of nonlinear ensemble transport filters and smoothers to subspaces (found via the ensemble or other means) should similarly improve the scalability of these algorithms in higher-dimensional systems.
    \item Likewise, \textbf{graph structure learning} methods \citep{Drton2017StructureModeling,Baptista2021LearningTransport}, for continuous non-Gaussian distributions, offer an approach to directly learn the conditional dependence structure of joint state and observation distributions relevant to smoothing. These constitute an alternative to the heuristics proposed in Section~\ref{subsec:L96}. Crucially, many such methods avoid constructing dense matrices or other objects that are of the size of the entire graph \citep{dong2022nonparametric}. Knowing this dependence structure permits the use of sparser transport maps, improving the sample efficiency of map learning (important for low ensemble sizes) and improving scalability to high dimensions.

    \item Finally, the \textbf{dynamic adaptation} of map parameterizations holds significant potential beyond the simple variations of polynomial degree that we explored here. In principle, it is possible to adjust the fine-grained complexity of individual map components $S_{k}$, including anisotropic dependence on arguments $x_{j}$, $j=1,\dots,k$ and their interactions. Recent work in this direction for (conditional) density estimation problems can be found, for example, in \citet{Baptista2020OnMaps}.
\end{itemize}
Progress along these directions will capitalize on the unique properties of ensemble transport methods, and help realize their potential for filtering and smoothing in the high-dimensional, low-ensemble-size systems prevalent across the environmental sciences and beyond. 

\section{Acknowledgements} \label{acknowledgements}

We thank the anonymous reviewers for their many thoughtful and helpful comments. The research of MR leading to these results has received funding from the Swiss National Science Foundation under the Early PostDoc Mobility grant P2NEP2 191663. RB and YM also acknowledge support from the US Department of Energy AEOLUS Mathematical Multifaceted Integrated Capabilities Center (MMICC) under award DE-SC0019303. MR and YM also acknowledge support from the Office of Naval Research Multidisciplinary University Research Initiative on Integrated Foundations of Sensing, Modeling, and Data Assimilation for Sea Ice Prediction under grant award N00014-20-1-2595.

\begin{appendices}
\section{Map optimization for separable map component functions}\label{sec:AppendixA}

Separable map component functions $S_{k}\colon\mathbb{R}^K\rightarrow\mathbb{R}$ follow the structure:
\begin{equation}
    S_{k}(w_{1},\dots,w_{k}) = u_{\text{off}}^{k}(w_1,\dots,w_{k-1}) + u_{\text{diag}}^{k}(w_{k}),
    \label{apeq:separable_basic}
\end{equation}
where $u_{\text{off}}^{k}$ contains all off-diagonal terms, namely those which do \textit{not} depend on the last input $w_{k}$, and $u_{\text{diag}}^{k}$ contains all monotone terms, that is those which \textit{only} depend on $w_{k}$. In this formulation, we thus assume that there are no cross-terms which depend on both $w_{k}$ and any $\w_{<k}$, such as $w_1w_{k}^{2}$. We call such map formulations \textit{separable in $w_{k}$}. 

Next, we assume the functions $u_{\text{off}}^{k}$ and $u_{\text{diag}}^{k}$ are linear in the coefficients, which is to say that have the form
\begin{align}
    u_{\text{off}}^{k}(w_{1},\dots,w_{k-1}) &= \sum_{j=1}^{E} \psi_{\text{off},j}(w_{1},\dots,w_{k-1})c_{\text{off},j} \label{eq:offdiag_expansion} \\
    u_{\text{diag}}^{k}(w_{k}) &= \sum_{j=1}^{D} \psi_{\text{diag},j} (w_{k})c_{\text{diag},j} \label{eq:diag_expansion},
\end{align}
where we have omitted the $k$ subscripts for ease of notation, and $(\psi_{\text{off},j})$ and $(\psi_{\text{diag},j})$ are basis functions depending on inputs variables $\w_{<k}$ and $w_k$, respectively, 
When the basis functions in Equations~\ref{eq:offdiag_expansion} and~\ref{eq:diag_expansion} are evaluated at $N$ samples, we can assemble the matrices $\boldsymbol{\Psi}_{\text{off}}^{k} \in \mathbb{R}^{N \times D}$ and $\boldsymbol{\Psi}_{\text{diag}}^{k} \in \mathbb{R}^{N \times E}$, and write the evaluations of $u_{\text{off}}$ and $u_{\text{diag}}$ concisely as $\boldsymbol{\Psi}_{\text{off}}\mathbf{c}_{\text{off}}$ and $\boldsymbol{\Psi}_{\text{diag}}\mathbf{c}_{\text{diag}}$, where $\mathbf{c}_{\text{off}}$ and $\mathbf{c}_{\text{diag}}$ are column vectors of length $D$ and $E$ containing the respective basis functions' coefficients. 

Using these evaluations in Equation~13, we can re-write the empirical objective function for the map component as
\begin{equation}
    \widehat{\mathcal{J}}_{k}(\mathbf{c}_{\text{off}},\mathbf{c}_{\text{diag}}) = \frac{1}{2N}\lVert\boldsymbol{\Psi}_{\text{off}}\mathbf{c}_{\text{off}} + \boldsymbol{\Psi}_{\text{diag}}\mathbf{c}_{\text{diag}}\rVert_{2}^{2} - \frac{1}{N}\sum_{i=1}^{N}\log \left|\partial_{k}\boldsymbol{\Psi}_{\text{diag}}\mathbf{c}_{\text{diag}}\right|,
\end{equation}
In practice, we also consider $L^2$ regularization by adding penalty terms to this objective function
\begin{equation}
    \widetilde{\mathcal{J}}_{k}(\mathbf{c}_{\text{off}},\mathbf{c}_{\text{diag}}) = \frac{1}{2N}\lVert\boldsymbol{\Psi}_{\text{off}}\mathbf{c}_{\text{off}} + \boldsymbol{\Psi}_{\text{diag}}\mathbf{c}_{\text{diag}}\rVert_{2}^{2} - \frac{1}{N}\sum_{i=1}^{N}\log \left|\partial_{k}\boldsymbol{\Psi}_{\text{diag}}\mathbf{c}_{\text{diag}}\right| + \lambda\lVert\mathbf{c}_{\text{off}}\rVert_{2}^{2} + \lambda\lVert\mathbf{c}_{\text{diag}}\rVert_{2}^{2},
    \label{apec:separable_objective_raw}
\end{equation}
where $\lambda > 0$ is a regularization parameter. We seek the variables $\mathbf{c}_{\text{off}}^{*}$ and $\mathbf{c}_{\text{diag}}^{*}$ which minimize this objective. 
Fortunately, we can simplify this optimization problem by finding the off-diagonal coefficients $\mathbf{c}_{\text{off}}$ which minimize the objective in closed form. For each setting of $\mathbf{c}_{\text{diag}}$, the gradient of the objective in Equation~\ref{apec:separable_objective_raw} with respect to the  $\mathbf{c}_{\text{off}}$ is
\begin{equation}
    \boldsymbol{\nabla}_{\mathbf{c}_{\text{off}}} \widetilde{\mathcal{J}}_k(\mathbf{c}_{\text{off}},\mathbf{c}_{\text{diag}}) = \frac{1}{N}(\boldsymbol{\Psi}_{\text{off}}\mathbf{c}_{\text{off}} + \boldsymbol{\Psi}_{\text{diag}}\mathbf{c}_{\text{diag}})^\top\boldsymbol{\Psi}_{\text{off}} + 2\lambda\mathbf{c}_{\text{off}}^{ \top}.
\end{equation}
By setting the gradient $\boldsymbol{\nabla}_{\mathbf{c}_{\text{off}}} \widetilde{\mathcal{J}}_k(\mathbf{c}_{\text{off}},\mathbf{c}_{\text{diag}})$ to zero, we obtain the optimal coefficients 
\begin{equation}
    \mathbf{c}_{\text{off}}^{*} = - (\boldsymbol{\Psi}_{\text{off}}^\top\boldsymbol{\Psi}_{\text{off}} + 2\lambda N \mathbf{I})^{-1} \boldsymbol{\Psi}_{\text{off}}^\top\boldsymbol{\Psi}_{\text{diag}}\mathbf{c}_{\text{diag}} = -\mathbf{A}\mathbf{c}_{\text{diag}},
    \label{apeq:c_off_opt}
\end{equation}
where $\mathbf{A} \coloneqq  (\boldsymbol{\Psi}_{\text{off}}^\top\boldsymbol{\Psi}_{\text{off}} +  2\lambda N \mathbf{I})^{-1} \boldsymbol{\Psi}_{\text{off}}^\top\boldsymbol{\Psi}_{\text{diag}} \in \R^{E \times D}$. Substituting the expression for $\mathbf{c}_{\text{off}}$ in Equation~\ref{apec:separable_objective_raw}, we obtain an objective which only depends on $\mathbf{c}_{\text{diag}}$. That is,
\begin{equation}
    \widetilde{\mathcal{J}}_{k}(\mathbf{c}_{\text{off}}^*,\mathbf{c}_{\text{diag}}) = \frac{1}{2N}\lVert(\boldsymbol{\Psi}_{\text{diag}} - \boldsymbol{\Psi}_{\text{off}}\mathbf{A} ) \mathbf{c}_{\text{diag}} \rVert_{2}^{2} - \frac{1}{N} \sum_{i=1}^{N}\log \left|\partial_{k}\boldsymbol{\Psi}_{\text{diag}}\mathbf{c}_{\text{diag}}\right| + \lambda\lVert \mathbf{A}\mathbf{c}_{\text{diag}}\rVert_{2}^{2} + \lambda\lVert\mathbf{c}_{\text{diag}}\rVert_{2}^{2}.
    \label{apec:separable_objective_simplified}
\end{equation}
Expanding all of the terms, we find:
\begin{equation}
\begin{aligned}
    \widetilde{\mathcal{J}}_{k}(\mathbf{c}_{\text{off}}^*,\mathbf{c}_{\text{diag}}) = &\mathbf{c}_{\text{diag}}^\top \left(\frac{1}{2N}(\boldsymbol{\Psi}_{\text{diag}} - \boldsymbol{\Psi}_{\text{off}}\mathbf{A} )^\top(\boldsymbol{\Psi}_{\text{diag}} - \boldsymbol{\Psi}_{\text{off}}\mathbf{A} ) + \lambda\mathbf{A}^\top\mathbf{A} + \lambda\mathbf{I} \right)\mathbf{c}_{\text{diag}} \\
    &- \frac{1}{N}\sum_{i=1}^{N}\log \left|\partial_{k}\boldsymbol{\Psi}_{\text{diag}}\mathbf{c}_{\text{diag}}\right|.
\end{aligned}
\end{equation}
Introducing a final substitution in terms of matrix $\mathbf{B} \in \R^{D \times D}$, we obtain:
\begin{equation}
    \begin{aligned}
    \widetilde{\mathcal{J}}_{k}(\mathbf{c}_\text{diag}) &=  \mathbf{c}_{\text{diag}}^\top\mathbf{B}\mathbf{c}_{\text{diag}} -  \sum_{i=1}^{N}\log \left|\partial_{k}\boldsymbol{\Psi}_{\text{diag}}\mathbf{c}_{\text{diag}}\right|, \text{  with} \\
    \mathbf{B} &= \frac{1}{2N}(\boldsymbol{\Psi}_{\text{diag}} - \boldsymbol{\Psi}_{\text{off}}\mathbf{A} )^\top(\boldsymbol{\Psi}_{\text{diag}} - \boldsymbol{\Psi}_{\text{off}}\mathbf{A} ) + \lambda(\mathbf{A}^\top\mathbf{A} + \mathbf{I}) \\
    \mathbf{A} &= (\boldsymbol{\Psi}_{\text{off}}^\top\boldsymbol{\Psi}_{\text{off}} + 2\lambda N \mathbf{I})^{-1} \boldsymbol{\Psi}_{\text{off}}^\top\boldsymbol{\Psi}_{\text{diag}},
    \end{aligned}
\label{apeq:separable_objective_complete}
\end{equation}
where we note that $\mathbf{A}$ and $\mathbf{B}$ only have to be calculated once for the optimization problem, as they are independent of the diagonal coefficients we wish to optimize for. We minimize Equation~\ref{apeq:separable_objective_complete} with a positivity constraint on $\mathbf{c}_{\text{diag}}$ to ensure monotonicity of $S_{k}$ with respect to $x_{k}$. After identifying the optimal coefficients $\mathbf{c}_{\text{diag}}^{*}$, we substitute
the solution in Equation~\ref{apeq:c_off_opt} to find the corresponding optimal off-diagonal coefficients  $\mathbf{c}_{\text{off}}^{*}$. 

\section{Map optimization for integrated component functions}\label{sec:AppendixB}

For integrated map components, we let the component function $S_{k}\colon\mathbb{R}^K\rightarrow\mathbb{R}$ follow the structure
\begin{equation}
    S_{k}(w_{1},\dots,w_{k}) = u_{\text{off}}^{k}(w_1,\dots,w_{k-1}) + \int_0^{w_k} r(u_{\text{diag}}^{k}(w_{1},\dots,\omega))\mathrm{d}\omega,
    \label{apeq:intrec_basic}
\end{equation}
where $u_{\text{off}}^{k}(w_{1},\dots,w_{k-1})$ contains all off-diagonal terms, namely those which do \textit{not} depend on the last dimension $w_{k}$, the term $u_{\text{diag}}^{k}(w_{1},\dots,w_{k})$ contains all terms which depend on the last dimension $w_{k}$, and $r\colon \mathbb{R} \rightarrow \mathbb{R}_{+}$ is a rectifier which takes arbitrary input and maps it to strictly positive values. Integrating the rectifier yields the monotone increasing function $S_{k}$ in $w_{k}$. To define the optimization objective, we substitute the expression above in the objective function for the map in Equation~13. Given $N$ target samples $\w^{i}$, the resulting objective function for the unconstrained functions $u_{\text{off}}^k$ and $u_{\text{diag}}^k$ is given by
\begin{equation}
    \begin{aligned}
    \widehat{\mathcal{J}}_{k}(u_{\text{off}}^k,u_{\text{diag}}^k) = \frac{1}{N}\sum_{i=1}^{N}&\frac{1}{2} \left(u_{\text{off}}^{k}(w_{1}^{i},\dots,w_{k-1}^{i}) + \int_0^{w_k^i} r(u_{\text{diag}}^{k}(w_{1}^{i},\dots,\omega))\partial \omega \right)^2 \\ &- \log \left|\partial_{w_k}\left(u_{\text{off}}^{k}(w_{1}^{i},\dots,w_{k-1}^{i}) + \int_0^{w_k^i} r(u_{\text{diag}}^{k}(w_{1}^{i},\dots,\omega)d\omega \right)\right|,
    \end{aligned}
\end{equation}
Applying the derivative in the second term removes the dependence on the off-diagonal map components and the integral:
\begin{equation}
    \begin{aligned}
    \widehat{\mathcal{J}}_{k}(u_{\text{off}}^k,u_{\text{diag}}^k) = &\frac{1}{N} \sum_{i=1}^{N}\frac{1}{2} \underbrace{\left(u_{\text{off}}^{k}(w_{1}^{i},\dots,w_{k-1}^{i}) + \int_0^{w_{k}^i} r(u_{\text{diag}}^{k}(w_{1}^{i},\dots,\omega) )\partial \omega \right)^2}_{S_{k}(w_{1}^{i},\dots,w_{k}^{i})^2} - \\
    &\log r(u_{\text{diag}}^{k}(w_{1}^{i}, \dots,w_{k}^{i})).      
    \end{aligned}
\end{equation}

\section{Map parameterization for the bimodal sine example}\label{sec:AppendixC}

This appendix describes the nonlinear maps used in the bimodal sine example in Section~5.1. Since we expect the conditional distributions for this problem to have two modes, the integrated representation described in Section~3.3.1 is a suitable choice for the map components. 

Furthermore, we anticipate the disjoint modes to have comparatively low spread with varying separation from each other. As a result, polynomial expansions are inefficient at 
recovering two narrow modes at varying distances because they require many
high-order terms. Using radial basis functions (RBFs) instead is substantially more efficient, since these basis functions can be positioned and scaled to capture multiple modes in the target distribution. In our implementation, we define the $i$th RBF with respect to $w_k$ as
\begin{equation}
    \operatorname{RBF}_{k,i}(w_{k}) = \frac{1}{\sqrt{2p\sigma_{k,i}^2}}\exp{\left(-\frac{(w_{k} - \mu_{k,i})^2}{2\sigma_{k,i}^2}\right)},
\end{equation}
for $l=1,\dots,L$. We assign the means of the $L$ RBFs $\mu_{k,i}$ along dimension $w_{k}$ according to the empirical quantiles of the marginal training samples, by setting each $\mu_{k,i}$ at the $i/(L+1)$ empirical quantile in $w_{k}$. For the scale factors, we use unit standard deviation $\sigma_{k,i}=1$. 

To formulate the lower map component $S_{2}$ using an exponential rectifier, we recall from Appendix~\ref{sec:AppendixB}:
\begin{equation}
    S_{2}(w_{1},w_{2}) = u_{\text{off}}^{2}(w_1) + \int_{0}^{w_2}\exp\left(u_{\text{diag}}^{2}(w_{1},\omega)\right)\partial \omega
\end{equation}
Since we expect two modes, we use two RBFs in each dimension $w_{k}$, and generate cross-terms by using the tensor product of these basis functions. The monotone and nonmonotone part of the map components are defined as follows:
\begin{equation}
    u_{\text{off}}^{2}(w_1) = c_{0} + c_{1}w_{1} + c_{2}\operatorname{RBF}_{1,1}(w_1) +  c_{3}\operatorname{RBF}_{1,2}(w_1),
\end{equation}
\begin{equation}
    \begin{aligned}
    & u_{\text{diag}}^{2}(w_{1},w_{2}) = & c_{4}\operatorname{RBF}_{1,1}(w_1)\operatorname{RBF}_{2,1}(w_2) \\  & & + c_{5}\operatorname{RBF}_{1,2}(w_1)\operatorname{RBF}_{2,1}(w_2) \\ & & + c_{6}\operatorname{RBF}_{1,1}(w_1)\operatorname{RBF}_{2,2}(w_2) \\ & & + c_{7}\operatorname{RBF}_{1,2}(w_1)\operatorname{RBF}_{2,2}(w_2)
    \end{aligned}
    \begin{aligned}
    \\ 
    \\
    \\
    .
    \end{aligned}
\end{equation}
This map structure is used for both the filter, for which we define $\w = \left[w_1,w_2\right]^\top := \left[y_t,x_t\right]^\top$, and the backward smoother, for which we define $\w = \left[w_1,w_2\right]^\top := \left[x_{s+1},x_{s}\right]^\top$. In the linear formulation of the filters and smoothers, 
the nonmonotone and monotone parts of the map are defined as:
\begin{equation}
    u_{\text{off}}^{2}(w_1) = c_{0} + c_{1}w_{1}
\end{equation}
\begin{equation}
    \begin{aligned}
    & u_{\text{diag}}^{2}(w_{2}) = c_{2}w_{2}.
    \end{aligned}
\end{equation}

\section{Bimodal experiment with true state-space equations}\label{sec:AppendixD}

The bimodal experiment in Section~5.1 considered a forecast model based on additive random noise. 
In this appendix, we will present results of linear and nonlinear filters and smoothers, for the case that the true state space equations are presumed known. For these experiments, we assume the initial condition for the two-dimensional state $\x_{1}=\left[x_{1}^{a},x_{1}^{b}\right]^\top$ is drawn from the bivariate Gaussian distribution:
\begin{equation}
    \x_{1}\sim\mathcal{N} \left (
    \begin{bmatrix*}
    0 \\
    0
    \end{bmatrix*},
    \begin{bmatrix*}
    0.2^{2} & 0 \\
    0 & 0.008^{2}
    \end{bmatrix*} \right ).
\end{equation}
We also consider three versions of this scenario here: One without forecast error, and two with additive forecast errors of varying magnitude sampled from the following bivariate Gaussian distributions:
\begin{equation}
\begin{aligned}
    \boldsymbol{\epsilon}_{\textrm{small}}&\sim\mathcal{N} \left (
    \begin{bmatrix*}
    0 \\
    0
    \end{bmatrix*},
    \begin{bmatrix*}
    0.01^{2} & 0 \\
    0 & 0.0004^{2}
    \end{bmatrix*} \right ) \\
    \boldsymbol{\epsilon}_{\textrm{large}}&\sim\mathcal{N} \left (
    \begin{bmatrix*}
    0 \\
    0
    \end{bmatrix*},
    \begin{bmatrix*}
    0.1^{2} & 0 \\
    0 & 0.004^{2}
    \end{bmatrix*} \right ).
\end{aligned}
\end{equation}
We use the same observation model as in Section~5.1. In addition, we also use minimal $L^2$-regularization for the nonlinear transport algorithms to prevent instabilities, with a constant regularization coefficient of $\lambda=10^{-6}$ for both the filter and smoother.

\begin{figure}
  \centering
  \includegraphics[width=\textwidth]{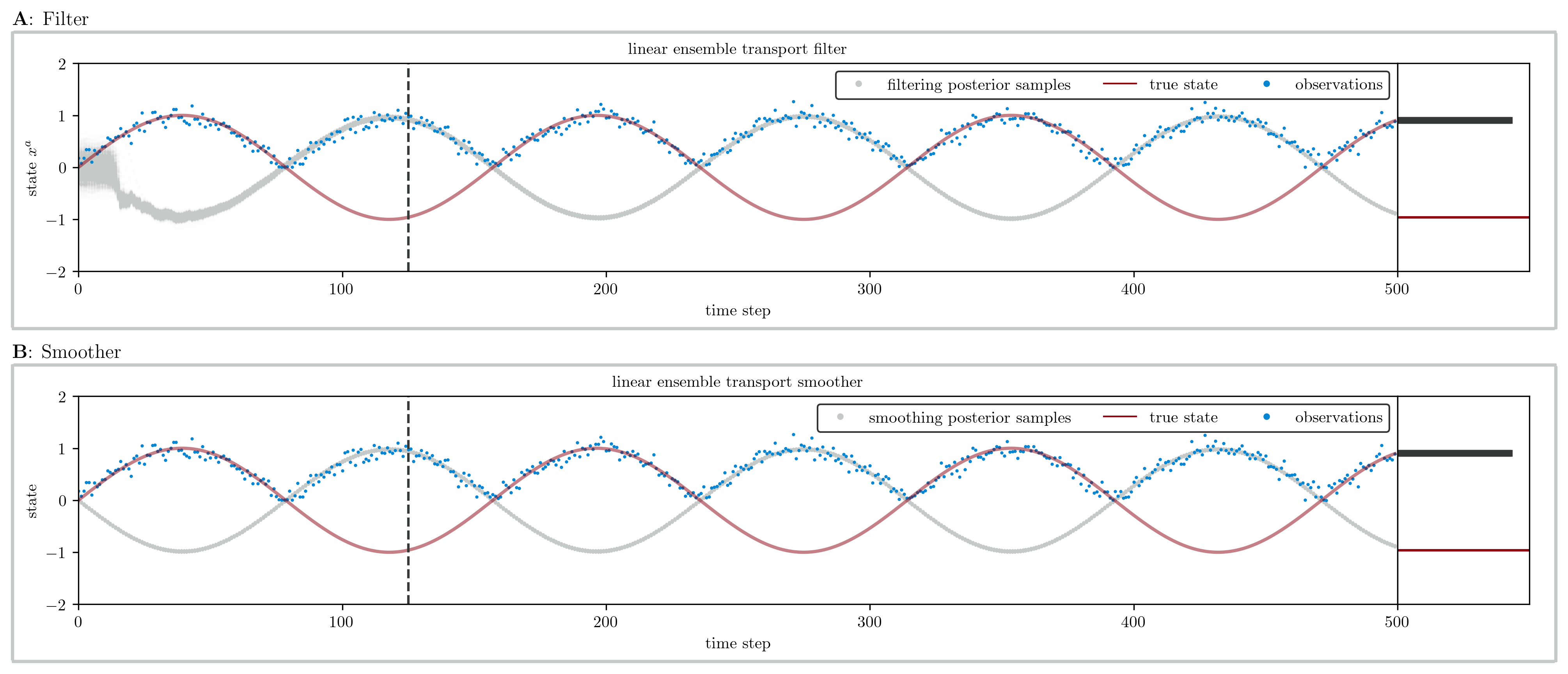}
  \caption{Scenario without forecast error: A combination of a filter (A) and a smoother (B) with linear updates fails to track multimodal distributions.}
  \label{fig:bimodal_sine_true_linear}
\includegraphics[width=\textwidth]{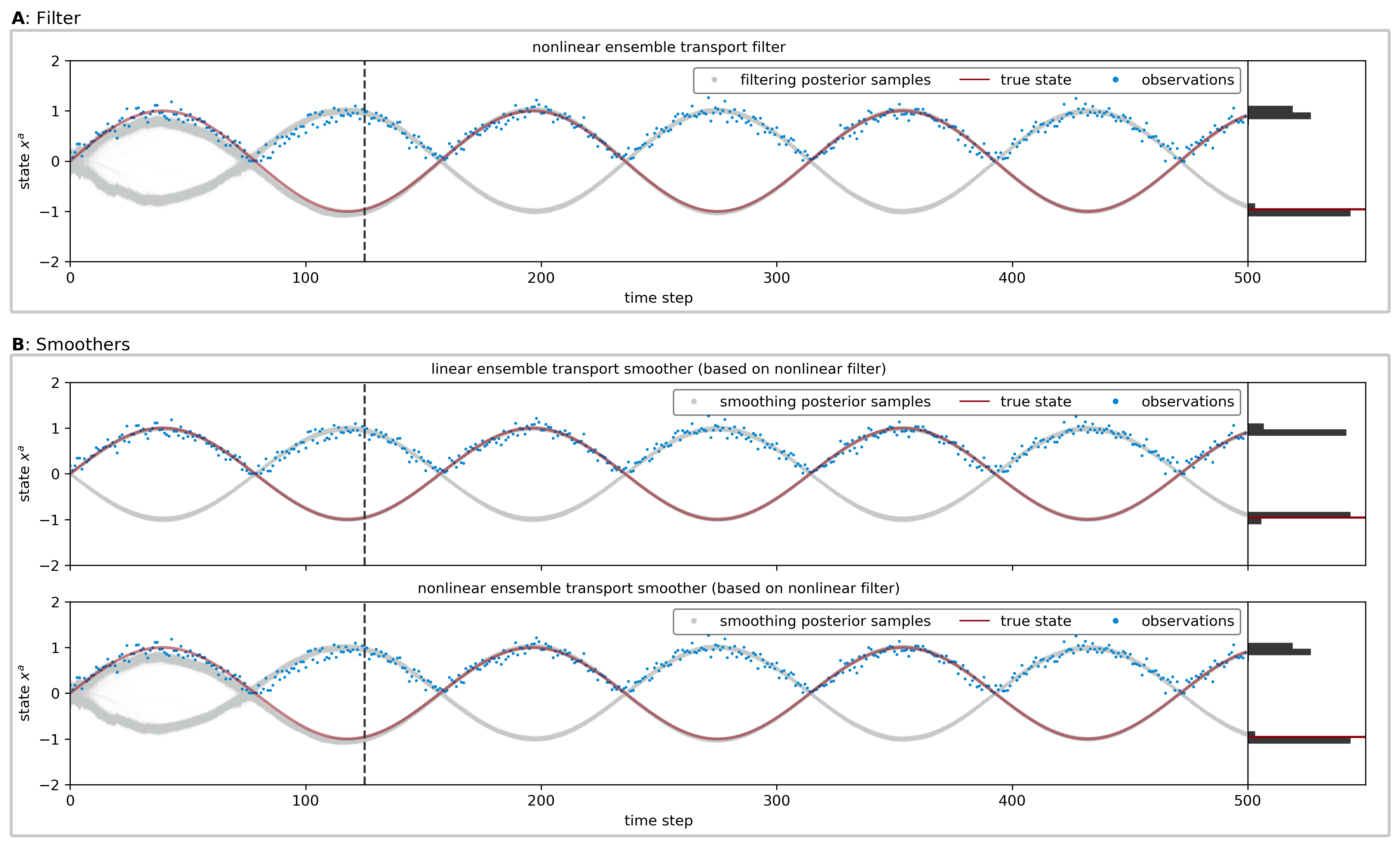}
  \caption{Scenario without forecast error: Filters with nonlinear updates can track multimodal distributions (A). In this setting, smoothers with both linear and nonlinear maps can preserve the bimodality (B).}
  \label{fig:bimodal_sine_true_nonlinear}
\end{figure}

For the scenario without forecast error, Figure~\ref{fig:bimodal_sine_true_linear} and Figure~\ref{fig:bimodal_sine_true_nonlinear} present results for the predicted state $x_{t}^{a}$ with both linear and non-linear updates, respectively. In the purely linear setting (Figure~\ref{fig:bimodal_sine_true_linear}), the ensemble collapses onto the branch corresponding to positive state values, and fails to capture the bimodal structure for the remainder of the time series. Similarly, the linear smoother remains confined to a single branch, but flattens out the early trajectories. Starting from a nonlinear filter (Figure~\ref{fig:bimodal_sine_true_nonlinear}) helps capturing the bimodality. In this case, the linear smoother retains this bimodal structure, and seems to provide narrower uncertainty bands than the nonlinear smoother. Recall that, opposed to the experiments in Section~5.1, this setting is slightly less challenging for the linear smoother because the ensemble remains separated in the second state $x_{t}^{b}$ even if it merges in $x_{t}^{a}$. In consequence, both the linear and nonlinear smoothers never have to create bimodality, only preserve it.

\begin{figure}
  \centering
  \includegraphics[width=\textwidth]{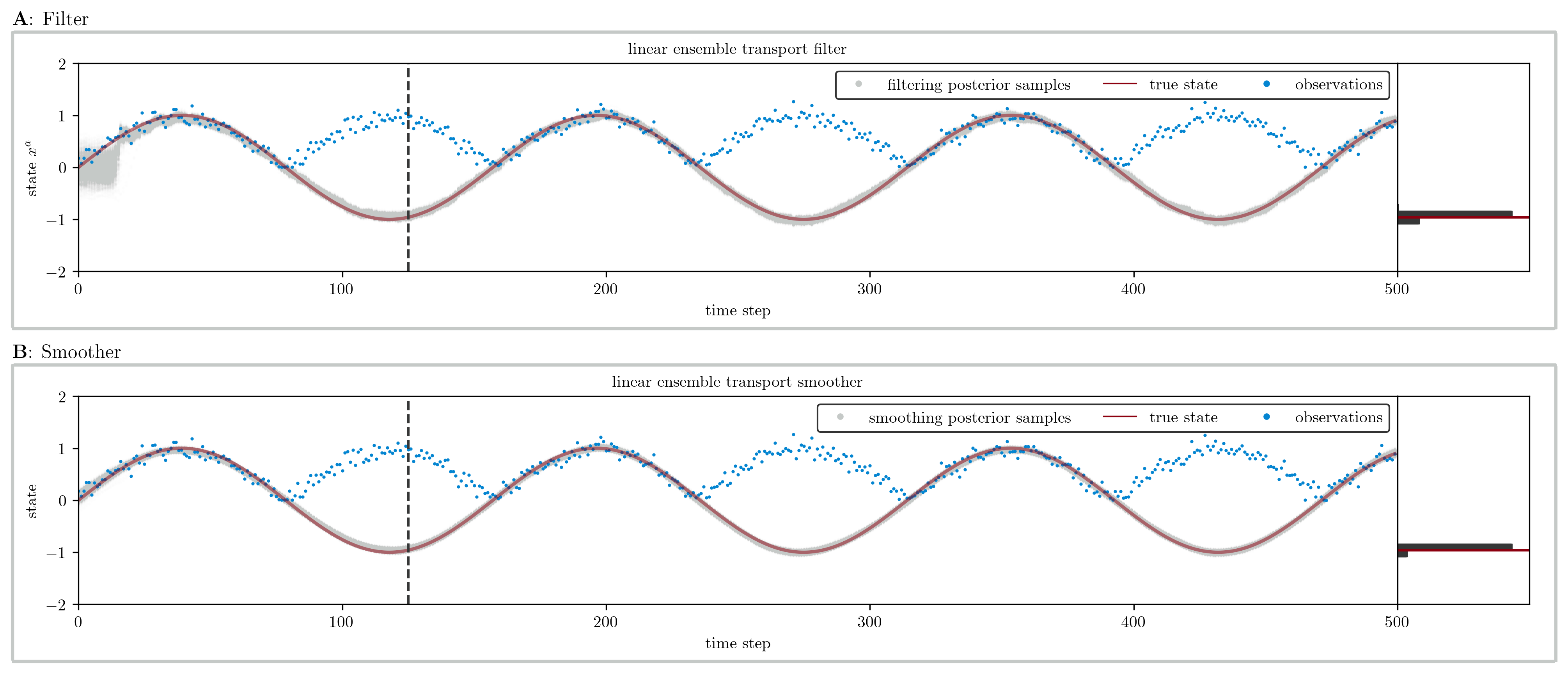}
  \caption{Scenario with small forecast error: A combination of a filter (A) and a smoother (B) with linear updates fails to track multimodal distributions.}
  \label{fig:bimodal_sine_true_error_linear}
\includegraphics[width=\textwidth]{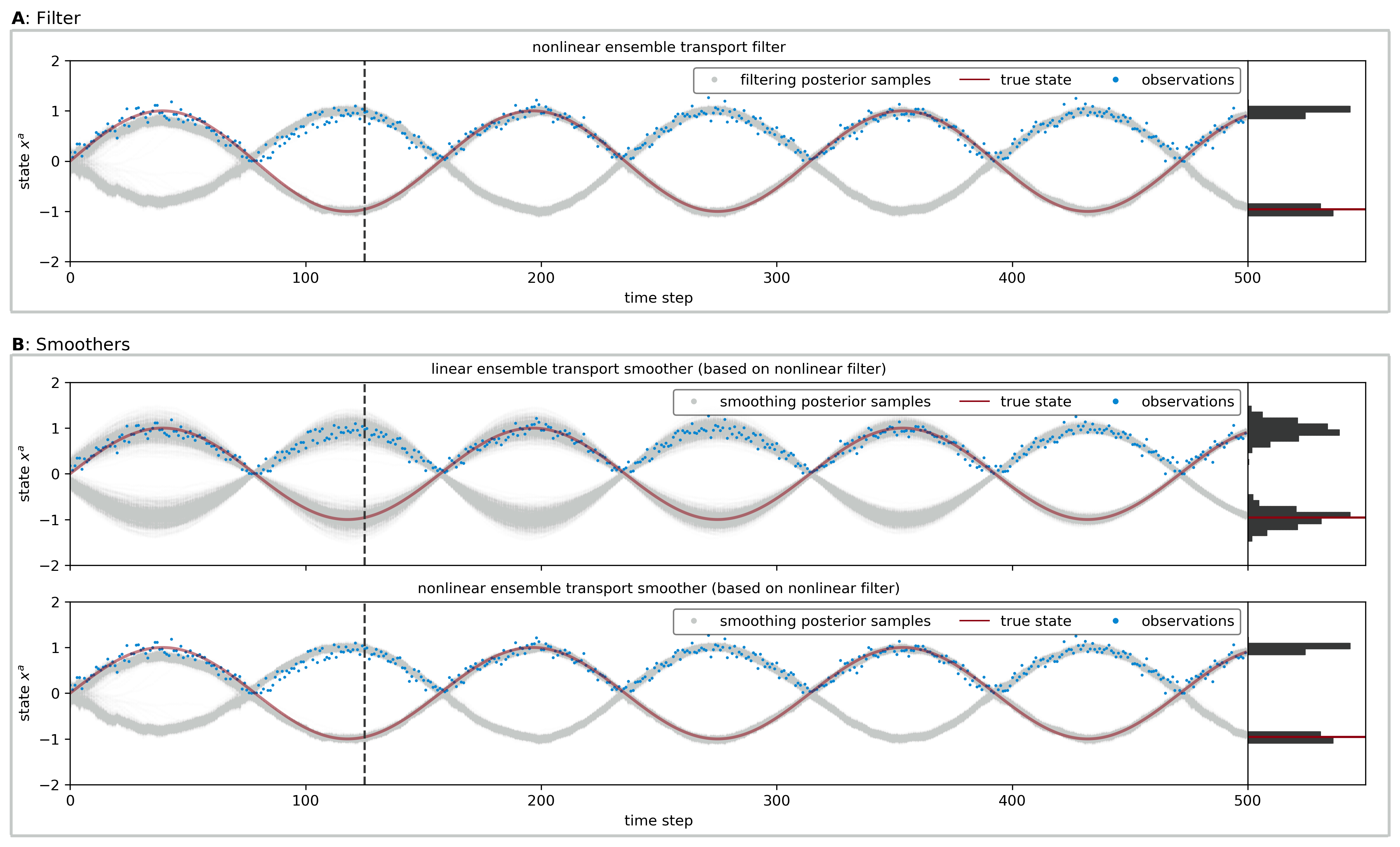}
  \caption{Scenario with small forecast error: Filters with nonlinear updates can track multimodal distributions (A). In this setting, smoothers with both linear and nonlinear maps can preserve the bimodality (B).}
  \label{fig:bimodal_sine_true_error_nonlinear}
\end{figure}

\begin{figure}
  \centering
  \includegraphics[width=\textwidth]{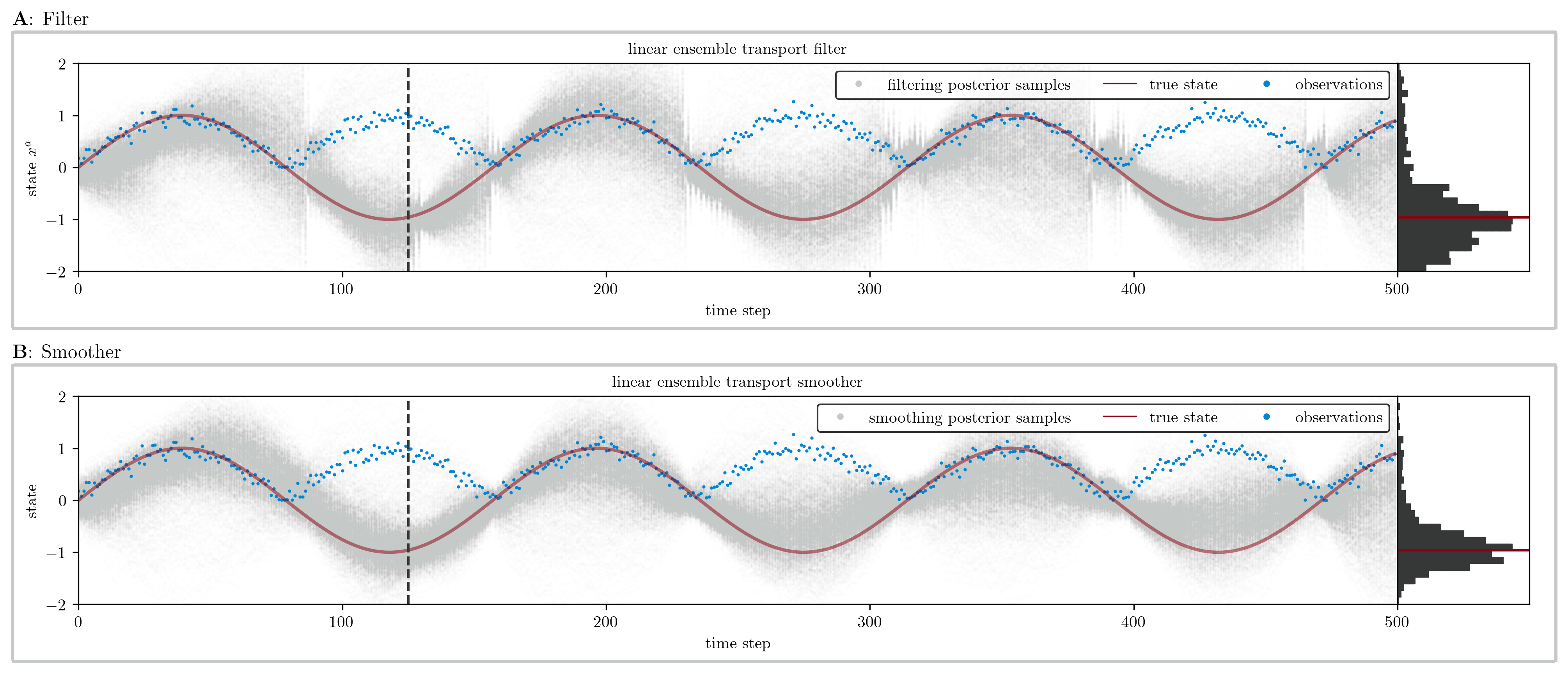}
  \caption{Scenario with large forecast error: A combination of a filter (A) and a smoother (B) with linear updates fails to track multimodal distributions.}
  \label{fig:bimodal_sine_true_high_error_linear}
\includegraphics[width=\textwidth]{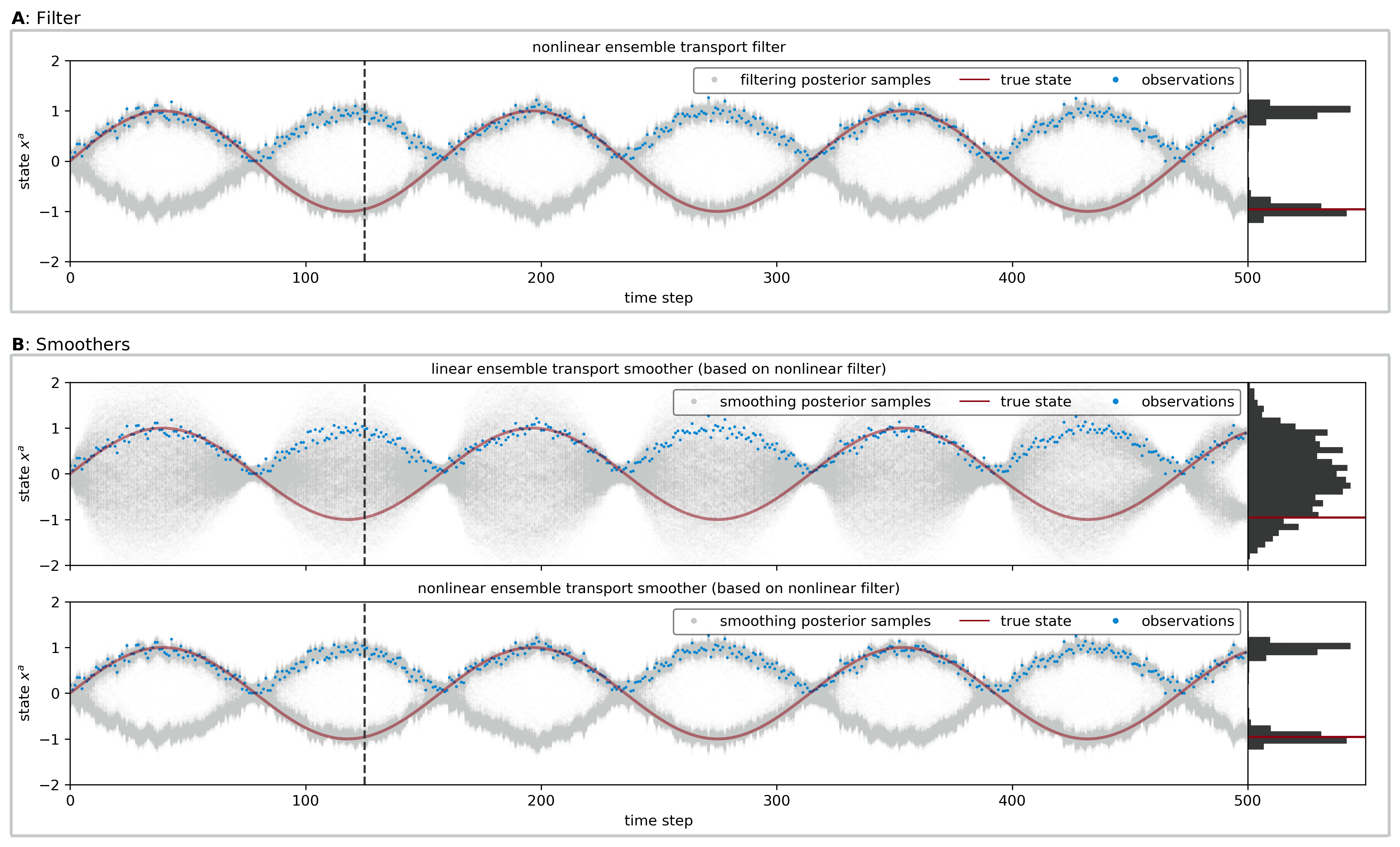}
  \caption{Scenario with large forecast error: Filters with nonlinear updates can track multimodal distributions (A). In presence of high error, a linear smoother cannot preserve the filter's bimodal structure, whereas a nonlinear smoother can (B).}
  \label{fig:bimodal_sine_true_high_error_nonlinear}
\end{figure}

In presence of small model error (Figure~\ref{fig:bimodal_sine_true_error_linear} and Figure~\ref{fig:bimodal_sine_true_error_nonlinear}), the purely linear smoother based on a nonlinear filter can still preserve the bimodal structure. However, the smoothing posterior begins to degrade during the backward pass (Figure~\ref{fig:bimodal_sine_true_error_nonlinear}B). This corruption is exacerbated by higher model errors (Figure~\ref{fig:bimodal_sine_true_high_error_linear} and Figure~\ref{fig:bimodal_sine_true_high_error_nonlinear}), where the linear backward smoother quickly blurs the smoothing posterior into a unimodal distribution, corrupting the bimodal structure (Figure~\ref{fig:bimodal_sine_true_high_error_nonlinear}B). This emphasizes that the linear backward smoother at best only preserves the non-Gaussianity, but cannot regenerate it. In the presence of model error, the nonlinear smoother thus reasserts its superior performance.

\section{Map parameterization for the Lorenz-63 and Lorenz-96 test cases} \label{sec:AppendixE}

This appendix describes the nonlinear map representations used in the chaotic systems in Sections~5.2 and~5.3. In both experiments, we use the separable map component functions described in Section~3.3.2.
We represent each map component $S_k$ as a sum of univariate functions $u_j \colon \R \rightarrow \R$ depending on each input argument $w_j$. That is,
\begin{equation}
    S_{k}(w_{1},\dots,w_k) = \sum_{j=1}^k u_j(w_j),
\end{equation}
where $u_k$ is defined to be monotone in $w_k$. We let the functions depending on $w_{j}$ for $j < k$ consist of an affine part (i.e., constant and linear monomials) followed by a linear combination of Hermite functions $\mathcal{H}_i(w)$ up to order $o > 1$, i.e, $$u_j(w_j) = c_0 + c_{1} w_j + \sum_{i=2}^{o} c_{i} \mathcal{H}_i(w_{j}), \quad j = 1,\dots,k-1.$$
Let us recall that a Hermite function $\mathcal{H}_{i} = H_{i}(w)\exp(-\frac{w^2}{4})$ of order $i$ comprises a probabilist Hermite polynomial $H_{i}$ of order $i$ multiplied with a Gaussian weight term. This weight term causes the tails of the Hermite function to revert to zero, preventing much of the instabilities found in the tails of high-order polynomials.

We let monotone function $u_k$ consist of strictly monotone basis functions. 
In this experiment, we use a linear combination of integrated RBFs (iRBF), with a left edge term (LET), and a right edge term (RET), i.e., 
\begin{equation} \label{eq:iRBF_edgeterms}
    u_k(w_k) = c_1 \operatorname{LET}(\Delta_{w_{k},1}) + \sum_{i=2}^{o-1} c_{i}\operatorname{iRBF}(\Delta_{w_{k},i}) + c_{o} \operatorname{RET}(\Delta_{w_{k},o}),
\end{equation}
with the basis functions defined as
\begin{equation}
    \begin{aligned}
    \operatorname{iRBF}(\Delta_{w,i}) & = \frac{1}{2}(1 + \operatorname{erf}(\Delta_{w,i})) \\        \operatorname{LET}(\Delta_{w,i}) & = \frac{1}{2}(\sqrt{2\pi}\Delta_{w,i}(1 - \operatorname{erf}(\Delta_{w,i})) - \sqrt{2/\pi}\exp(-\Delta_{w,i}^2)) \\        \operatorname{RET}(\Delta_{w,i}) & = \frac{1}{2}(\sqrt{2\pi}\Delta_{w,i}(1 + \operatorname{erf}(\Delta_{w,i})) + \sqrt{2/\pi}\exp(-\Delta_{w,i}^2)),
    \end{aligned}
    \label{apeq:RBF_terms}
\end{equation}
where the input variables $w$ are defined in terms of local coordinates:
\begin{equation}
    \Delta_{w,i} = \frac{w - \mu_{i,k}}{\sqrt{2\pi}\sigma_{i,k}}.
    \label{apeq:RBF_local_coordinates}
\end{equation}
$\operatorname{iRBF}$s are normalized error functions, inducing a smooth step centered at $\mu_{i,k}$. $\operatorname{LET}$s and $\operatorname{RET}$s define the left and right tails, respectively, providing linear tails on one side while tapering off to the other.
Similar to the bimodal example, the $L$ basis function along each dimension $w_{k}$ are placed at specific empirical quantiles of the training samples, with each $\mu_{i,k}$ placed at the $l/(L+1)$ empirical quantile in $w_{k}$. 
For order $o=1$, the monotone terms in Equation~\ref{eq:iRBF_edgeterms} are replaced by terms with only affine dependence in $w_{k}$.

\section{Iterative EnKS setup and results}\label{sec:AppendixF}

The reported iEnKS results for the Lorenz-63 in Section~5.2 and Lorenz-96 in Section~5.3 are obtained with the following procedure: 
\begin{enumerate}
    \item Use the Dapper toolbox to run iEnKS simulations for all combinations of ensemble size $N$, inflation factors $\gamma$, lag lengths $L$, and random seeds $r$. We save the results for the ensemble mean RMSE, the CRPS, and the average number of iEnKS iterations.
    \item Average all results across the random seeds for each combination of the remaining variables ($N$, $\gamma$, and $L$).
    \item Select the parameter $\gamma$ with the lowest average RMSE for every combination of $N$ and $L$
    \item Calculate the number of model evaluations for each combination of $(N,L)$ as $N\cdot(\operatorname{nIter}\cdot L+1)$, where $\operatorname{nIter}$ is the average number of iEnKS iterations for this combination.
    \item Finally, we determine the convex hull of the scatter cloud of model evaluations against the metric of interest (RMSE or CRPS), and then extract the lower edge of this hull. The resulting lower edges are reported in Figure~6.
\end{enumerate}

\begin{figure}
  \centering
  \includegraphics[width=\textwidth]{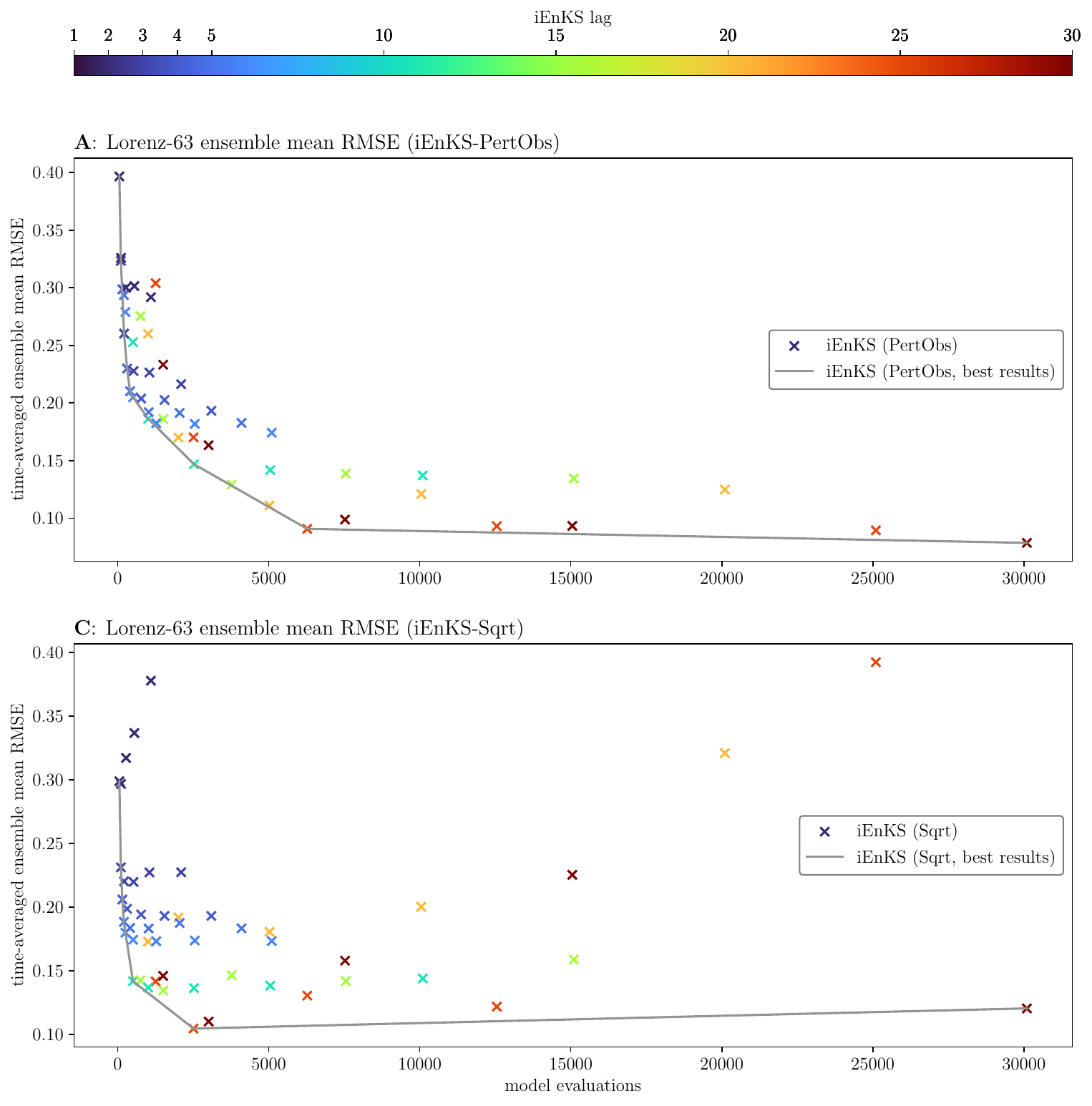}
  \caption{RMSE for the iEnKS-PertObs (A) and the iEnKS-Sqrt (B) against the number of model evaluations for the iEnKS simulations. The scatter color denotes Lag. The grey line delineates the lower edge of the convex hull of these simulations, representing the optimal results in terms of RMSE obtained in this study.}
  \label{apfig:iEnKS_L63}
\end{figure}

The scatter plots underlying the latter step are provided below for Lorenz-63 in Figure~\ref{apfig:iEnKS_L63}.

\end{appendices}

\FloatBarrier

\selectlanguage{english}
\bibliographystyle{plainnat}
\bibliography{references}

\end{document}